# Biphoton shaping with cascaded entangled-photon sources

Arash Riazi[1], Changjia Chen[1], Eric Y. Zhu[1], Alexey V. Gladyshev[2], Peter G. Kazansky[3], J. E. Sipe[4], L. Qian[1]

**Quantum entanglement is an integral part of quantum optics and has been exploited in areas such as computation, cryptography and metrology. The entanglement between photons can be present in various degrees of freedom (DOFs), and even the simplest bi-partite systems can occupy a large Hilbert space. Therefore, it is desirable to exploit this multi-dimensional space for various quantum applications by fully controlling the properties of the entangled photons in multiple DOFs. While current entangled-photon sources are capable of generating entanglement in one or more DOFs, there is currently a lack of practical techniques that can shape and control the entanglement properties in multiple DOFs. Here we show that cascading two or more entangled-photon sources with tunable linear media in between allows us to generate photon-pairs whose entanglement properties can be tailored and shaped in the frequency and polarisation domains. We first develop a quantum mechanical model to study the quantum state generated from the cascade structure with special considerations paid to the effects of pump temporal coherence, linear dispersion, and in-structure polarisation transformation applied between the entangled-photon sources. We then experimentally generate photon-pairs with tunable entanglement properties by manipulating the dispersion and birefringence properties of the linear medium placed in between two entangled-photon sources. This is done in an all-fibre, phase stable, and alignment-free configuration. Our results show that the cascade structure offers a great deal of flexibility in tuning the properties of entangled photons in multiple DOFs, opening up a new avenue in engineering quantum light sources.**

## Introduction

Entanglement is an essential resource in quantum optics and can be exploited for quantum information processing[1-5] and the study of fundamental physics[6-8]. New developments in quantum optics aim to generate entangled photons whose properties in various DOFs can be tailored and controlled. Frequency and polarisation of photons are robust DOFs often used in practical applications. As a result, a vast number of protocols and platforms have already been developed to exploit these two DOFs; for example, the spectrum of entangled photons (biphotons) has been exploited for scalable quantum information processing[1-2,9-11] and large alphabet quantum key distribution[12]; the ability to generate various biphoton polarisation states has also been recognized as a useful resource[13-16] for tests of local realism[6-7] and complementarity in physics[8,17].

In light of this, we can envision that the ability to tailor and shape the entanglement properties of biphotons in both frequency and polarisation DOFs would allow us to increase the amount of information that can be encoded into a biphoton state[5], enabling a variety of new applications in quantum optics. In order to achieve this goal, we first need spectral and polarisation shaping techniques for biphotons that are compatible with each other; these techniques should also be implementable in an integrated and scalable fashion.

So far, various techniques of biphoton spectral shaping have been demonstrated with spatial light modulators[18-19], spectral filtering[20-22], and tailoring the phase-matching structure of the nonlinear medium itself[23-24]. However, these techniques either introduce undesirable loss due to coupling and filtering[20-22], or impose considerable complications in the precise fabrication of the nonlinear structure[23-24]. Additionally, some of these techniques[18-19] cannot yet be integrated with waveguide-based biphoton sources, and therefore cannot take advantage of the greater mode confinement.

Various techniques to shape the polarization state of biphotons have also been demonstrated, typically through a combination of biphoton interference[14-16], unitary polarisation transformation[14-15], decoherence[14,16], and spatial mode selection[16]. However, these techniques have all been implemented using free-space setups and cannot be integrated with waveguide-based biphoton sources in a single platform. Moreover, precise beam alignment, spatial filtering, and phase stabilization are required for these techniques, which make them difficult to implement in integrated photonics. Finally, these techniques have not been shown to be simultaneously compatible with spectral tailoring. In fact, no practical approach to shape the biphotons simultaneously in both the spectral and polarization domains has been demonstrated.

In this paper, we demonstrate a technique that can shape biphoton states in both the frequency and polarisation domains by cascading two fibre-based entangled-photon sources[25-26] with a linear medium placed in between, which we refer to as the *middle section* (see **Fig. 1a**). Our cascade structure, which is essentially a nonlinear interferometer[27], can be pumped either with a long- or short-coherence-time laser, with each option providing a specific functionality for shaping the properties of biphotons. The spectrum and polarisation state of the biphotons generated from the cascade structure can be tailored by altering the dispersion and birefringence of the linear middle section. The all-fibre common-path configuration used here eliminates major issues in biphoton shaping, such as the requirement for beam alignment, coupling and filtering loss, and phase stabilization. More importantly, spectral and polarisation shaping techniques are now compatible with each other and can be simultaneously implemented in such a structure.

It is worth mentioning that the cascade structure we use here belongs to a more general class known as SU(1,1) nonlinear interferometer[27-28]. The high-gain regime of these interferometers has been extensively studied[27-31] and utilized[32-34] to obtain the Heisenberg limit in phase measurement. On the other hand, the spontaneous regime of these interferometers has also been studied both theoretically[28,35-37], and experimentally[38-41] to investigate more abstract concepts such as

[1]Dept. of Electrical and Computer Engineering, University of Toronto, 10 King's College Rd., Toronto, Canada M5S 3G4; [2]Fibre Optics Research Center, Russian Academy of Sciences, 38 Vavilov Street, 119333 Moscow, Russia; [3]Optoelectronics Research Centre, University of Southampton, Southampton SO17 1BJ, United Kingdom; [4]Dept. of Physics, University of Toronto, 60 St. George St., Toronto, Canada M5S 1A7
arash.riazi@mail.utoronto.ca

"induced coherence" effect[38,42-43]. These studies have since found their applications in measuring absorption[44], refractive index[45], and dispersion[46] of linear media.

Our work, however, differs from previous studies in that we utilize cascaded biphoton sources (a nonlinear interferometer) in two DOFs (frequency and polarisation) to generate biphotons with tunable entanglement properties in spectral and polarisation domains. Furthermore, the quantum mechanical treatment of the cascade structure we present here is more comprehensive as it takes into account the collective effects of the pump temporal coherence, the chromatic dispersion in the structure, and the polarisation transformations on the biphoton state generated from the cascade structure. Finally, our formulation can be generalized to other waveguide-based entangled-photon sources, including those in integrated photonics.

The organization of the paper is as follows: We first present a quantum mechanical model for the biphoton state at the output of the cascade structure, taking into account (1) the temporal coherence of the pump (referred to below simply as the pump coherence), (2) the dispersion properties of both linear and nonlinear segments, and (3) the polarisation transformation applied in the linear middle section of the cascaded structure. We then use our model to study the spectrum and polarisation state of the biphotons under various pump coherence conditions and polarisation transformations caused by the middle (linear) section. Finally, using two periodically-poled silica fibres[26] (PPSFs) as biphoton sources, we experimentally demonstrate: (1) the ability to generate biphotons with modified spectra, for various pump coherence conditions and the linear properties of the middle section; and (2) the ability to generate various biphoton polarisation states with properties such as tunable degree of polarisation entanglement.

## Results

**The cascade structure and theoretical framework**

The general two-segment cascade structure is shown in **Fig. 1a**. Two identical second-order nonlinear segments are connected via a "middle section" consisting of a linear optical medium, and an inline polarisation controller (PC). By using the nonlinear waveguide scattering theory presented earlier[47-48], we model biphoton generation in our cascade structure. For our formulation, we consider only *type-II* SPDC phase-matching; however, it can be trivially generalized to other SPDC phase-matchings as well. We define horizontal (*H*), and vertical (*V*) polarisations according to the principal axes (polarisation eigenmodes) of the nonlinear segments. Note that due to the polarisation transformation in the middle section, light polarized along one of the principal axes in the first nonlinear segment will not generally be polarized along the same principal axes of the second nonlinear segment.

We model the pump in **Fig. 1a** as a quasi-monochromatic field with a finite coherence time of $\tau_C$. In our model, the pump field is a succession of coherent packets[49-50] (see **Fig. 1b**) in which the electric field oscillates with a constant angular frequency of $\bar{\omega}_p$ (see **supplement, section 2**); the initial phase of the electric field within each packet is assumed to be constant, however, it is statistically distributed for each packet[49-50]. The generalized creation operator for the $\mathcal{L}^{th}$ pump packet with polarisation $S$ is denoted by $\hat{A}^\dagger_{\mathcal{L},S} = \int dk_P f^*_\mathcal{L}(k_P) \hat{a}^\dagger_{PSk_P}$, where $f_\mathcal{L}(k_P)$ includes the spectral behavior of that packet (see **supplement, section 2**) and is normalized according to $\int dk_P |f_\mathcal{L}(k_P)|^2 = 1$. The quantum state of each pump packet incident on the structure is taken to be a coherent state in vertical polarisation and can be written as $|\alpha_{\mathcal{L},V}\rangle = e^{\alpha_\mathcal{L} \hat{A}^\dagger_{\mathcal{L},V} - h.c.}|vac\rangle$, where $|\alpha_\mathcal{L}|^2$ is the average photon number inside the $\mathcal{L}^{th}$ pump packet. Since the field operators of different packets commute due to $\left[\hat{A}_{\mathcal{L},V}, \hat{A}^\dagger_{\mathcal{L}',V}\right] = \delta_{\mathcal{L},\mathcal{L}'}$ (see

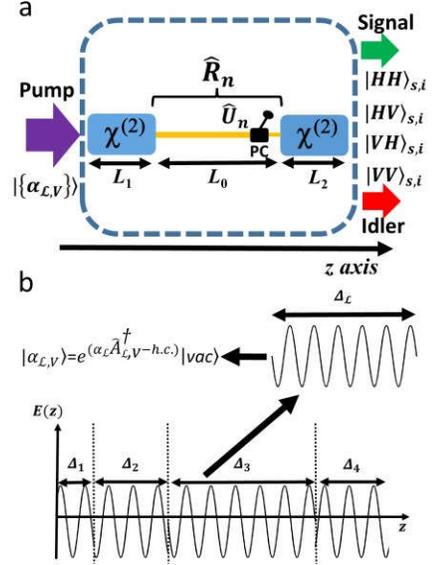

**Figure 1** (a) A two-segment cascade structure made up of two second-order nonlinear media of the lengths $L_1$ and $L_2$; the pump coherent packets with $V$ polarisation, $|\{\alpha_{\mathcal{L},V}\}\rangle$, enter the structure at $z = 0$. Depending on the transformation in the middle section $\hat{R}_n$, the polarisation states of the downconverted photon pairs at the output of the cascade structure could be a superposition of all four states ($|HH\rangle_{\omega_A,\omega_B}$, $|HV\rangle_{\omega_A,\omega_B}$, $|VH\rangle_{\omega_A,\omega_B}$, $|VV\rangle_{\omega_A,\omega_B}$). (b) The quasi-monochromatic model of the pump field[49]; the pump consists of coherent packets, each of which has a length of $\Delta_\mathcal{L}$. The initial phase of the electric field within each packet changes randomly from packet to packet[50].

**supplement, section 2**), we can write down the quantum state of the pump at the input of the cascade structure as:

$$|\{\alpha_{\mathcal{L},V}\}\rangle = e^{\Sigma_\mathcal{L}(\alpha_\mathcal{L} \hat{A}^\dagger_{\mathcal{L},V} - h.c.)}|vac\rangle. \quad (1)$$

In the weak conversion limit with negligible probability of multi-pair generation, the quantum state of the down-converted light for an individual nonlinear segment with a length of $L_1$ can be described as:

$$|\Psi_{seg}\rangle = |vac\rangle + \sum_\mathcal{L} \alpha_\mathcal{L} \int d\omega_P d\omega_A d\omega_B \mathcal{A}_0^{type\,II}(\omega_P, \omega_A, \omega_B) f_\mathcal{L}(\omega_P) \delta(\omega_P - \omega_A - \omega_B) \times$$
$$\left[L_1 \text{sinc}\left(\frac{\Delta k^{(1)}_{VHV} L_1}{2}\right) \hat{a}^\dagger_{AH} \hat{a}^\dagger_{BV} + e^{i\Lambda(\omega_A,\omega_B)L_1} \text{sinc}\left(\frac{\Delta k^{(1)}_{VVH} L_1}{2}\right) \hat{a}^\dagger_{AV} \hat{a}^\dagger_{BH}\right]|vac\rangle_{\omega_A,\omega_B}, \quad (2)$$

where the sum is over all the pump packets, and $\omega_P$, $\omega_A$, and $\omega_B$ are the angular frequencies of the pump, signal, and idler fields, respectively; $\hat{a}^\dagger_{AS'}$ ($\hat{a}^\dagger_{BS''}$) is the creation operator of the signal (idler) mode with $S'(S'')$ polarisation; the quantity $\mathcal{A}_0(\omega_P, \omega_A, \omega_B)$ includes the nonlinear susceptibilities and other phase factors (see **supplement, section 5**); $\Delta k^{(m)}_{S\,S'\,S''} = k^{(m)}_{PS}(\omega_P) - k^{(m)}_{AS'}(\omega_A) - k^{(m)}_{BS''}(\omega_B) - k_{QPM}$, where the first and the second subscripts of the wavenumbers refer to the field and its polarisation, respectively, while the superscript *m* refers to the nonlinear media; $k_{QPM}$ is the quasi-phase-matching wavenumber of the nonlinear medium. Note that $\Delta k^{(1)}_{VHV} L_1$ [**Eq. (2)**] in general is different from $\Delta k^{(1)}_{VVH} L_1$; however, due to the slowly-varying nature of sinc function [in comparison with other phase factors in **Eq. (2)**] and small difference between $\Delta k^{(1)}_{VHV} L_1$ and $\Delta k^{(1)}_{VVH} L_1$ over the phase-matching (full width at half maximum) bandwidth of the signal and idler, we have $\text{sinc}(\frac{\Delta k^{(1)}_{VHV} L_1}{2}) \approx \text{sinc}(\frac{\Delta k^{(1)}_{VVH} L_1}{2})$. Finally, the quantity $\Lambda(\omega_A, \omega_B)$ [related to group birefringence, see ref. 51] is defined as:

$$\Lambda(\omega_A, \omega_B) = [k_{AV}^{(1)}(\omega_A) - k_{AH}^{(1)}(\omega_A) + k_{BH}^{(1)}(\omega_B) - k_{BV}^{(1)}(\omega_B)]\frac{L_1}{2}. \quad (3)$$

Henceforth, we shall drop the angular frequency notation for the wavenumber $k(\omega)$; for $\mathcal{A}_0^{type\,II}(\omega_P, \omega_A, \omega_B)$ and $\delta(\omega_P - \omega_A - \omega_B)$, we simply write $\mathcal{A}_0^{type\,II}$, and $\delta$, respectively.

Note that the SPDC emission within the PPSF [see **Eq. (2)**] results in both signal and idler photons traveling in one spatial mode. However, we can distinguish the two based on their angular frequencies; photons with angular frequencies greater than $\frac{\bar{\omega}_p}{2}$ are called signal; otherwise, they are called idler. Note that our definition of signal and idler best describes cases with narrowband pump [*e.g.* continuous (cw) pump].

We also remark that for most second-order nonlinear media, the biphoton state generated from type-II SPDC is not polarisation-entangled due to the walk-off caused by the frequency-dependent factor $e^{i\Lambda(\omega_A, \omega_B)}$ in **Eq. (2)**[see ref. 51]; however, because of the unique dispersive properties of poled-fibre[$\Lambda(\omega_A, \omega_B) \ll 1$, see ref. 51], type-II SPDC in PPSFs allows for the direct generation of polarisation-entangled photon-pairs[26,51-52]. Since we are using PSSF as our nonlinear medium, whenever type-II SPDC is involved, the biphoton state is polarisation-entangled.

**The state of the generated biphotons in the cascade structure**

Now we consider a cascade of two identical nonlinear segments pumped for type-II SPDC; the two nonlinear segments are connected via a linear medium (with a length of $L_0$), by which we shape the spectrum and polarisation state of the biphotons generated from the cascade structure. We derive the quantum state of the biphotons by employing several assumptions: (1) The collective transformation of the middle section in Jones space can be modeled by two consecutive transformations: A phase accumulation $e^{ik_n^{(0)}L_0}\hat{I}$, where $\hat{I}$ is a 2×2 identity matrix, and a unitary polarisation transformation $\widehat{U}_n = \begin{pmatrix} U_{1n} & U_{2n} \\ U_{3n} & U_{4n} \end{pmatrix}$[see ref. 53], where the subscripts $n = P, A$, and $B$. Accordingly, the collective transformation of the middle section (see **Fig. 1a**) becomes: $\hat{R}_n = e^{ik_n^{(0)}L_0}\begin{pmatrix} U_{1n} & U_{2n} \\ U_{3n} & U_{4n} \end{pmatrix}$(see **supplement, section 3**); (2) The middle section is assumed to have a weak wavelength-dependent birefringence such that $\hat{R}_A = \hat{R}_B \neq \hat{R}_P$. In other words, the signal and idler are assumed to undergo the same polarisation transformation, while the pump does not necessarily do so; (3) While the presence of the middle section may result in the pump polarisation having both *H* and *V* components when entering the second nonlinear segment, due to the phase-matching constraint (wavelength) for type-II SPDC, the *H* component of the pump will not contribute to other SPDC types (such as type-0) in the second segment (see **supplement, section 5.1**). So the effect of the middle section is to merely transform the polarisation state of the biphotons that could be generated in the first segment.

Under these assumptions, the quantum state of the biphotons at the output of the cascade structure can be written as a linear superposition of all possible biphoton states,

$$|\psi_{\text{CAS}}\rangle = |vac\rangle_{\omega_A, \omega_B} + \sum_{\mathcal{L}}\left\{\int d\omega_A d\omega_B\, \phi_{\mathcal{L},HH}(\omega_A, \omega_B)\hat{a}_{AH}^\dagger\hat{a}_{BH}^\dagger + \int d\omega_A d\omega_B\, \phi_{\mathcal{L},HV}(\omega_A, \omega_B)\hat{a}_{AH}^\dagger\hat{a}_{BV}^\dagger + \int d\omega_A d\omega_B\, \phi_{\mathcal{L},VH}(\omega_A, \omega_B)\hat{a}_{AV}^\dagger\hat{a}_{BH}^\dagger \quad (4) \\ \int d\omega_A d\omega_B\, \phi_{\mathcal{L},VV}(\omega_A, \omega_B)\hat{a}_{AV}^\dagger\hat{a}_{BV}^\dagger\right\}|vac\rangle_{\omega_A, \omega_B},$$

where $\phi_{\mathcal{L},S'S''}(\omega_A, \omega_B)$ is the biphoton wavefunction, corresponding to the $\mathcal{L}^{th}$ pump packet, which can be determined by the Hamiltonian treatment of the cascade structure (see **supplement, section 4-6**). Henceforth neglecting the vacuum contribution, we can write the quantum state of the biphotons generated from the cascade structure as:

$$|\psi_{\text{CAS}}\rangle = \sum_{\mathcal{L}}\alpha_{\mathcal{L}}\int d\omega_P d\omega_A d\omega_B\, \mathcal{A}_0^{type\,II} f_{\mathcal{L}}(\omega_P)L_1 \text{sinc}\left(\frac{\Delta k_{VHV}^{(1)}L_1}{2}\right)\delta \times \\
\{-e^{i\Gamma_i}[(U_{4A}U_{3B})^* + e^{i\Lambda}(U_{4B}U_{3A})^*]|HH\rangle_{\omega_A, \omega_B} \\
+ [(U_{4A}U_{1B})^* + e^{i\Lambda}(U_{3A}U_{2B})^* + U_{4p}e^{i(\Delta k^{(0)}L_0 + \Delta k_{VHV}^{(1)}L_1)}]|HV\rangle_{\omega_A, \omega_B} \\
+ e^{2i\Lambda}[(U_{2A}U_{3B})^* + e^{i\Lambda}(U_{1A}U_{4B})^* + e^{-i\Lambda}U_{4p}e^{i(\Delta k^{(0)}L_0 + \Delta k_{VHV}^{(1)}L_1)}]|VH\rangle_{\omega_A, \omega_B} \quad (5) \\
- e^{i\Gamma_s}[(U_{2A}U_{1B})^* + e^{i\Lambda}(U_{2B}U_{1A})^*]|VV\rangle_{\omega_A, \omega_B}\},$$

where we assumed the two nonlinear media have the same length ($L_1 = L_2$) and identical dispersion properties [*i.e.* $k_{H(V)}^{(1)}(\omega) = k_{H(V)}^{(2)}(\omega)$]; here $\Gamma_{A(B)} = (k_{A(B)V}^{(1)} - k_{A(B)H}^{(1)})L_1$, and $\Delta k^{(0)}L_0 = (k_P^{(0)} - k_A^{(0)} - k_B^{(0)})L_0$ is the phase introduced by the middle section. Note that due to the polarisation transformation in the middle section, $\phi_{\mathcal{L},HH}(\omega_A, \omega_B)$ and $\phi_{\mathcal{L},VV}(\omega_A, \omega_B)$ are now nonzero and the extra biphoton polarisation states $|HH\rangle_{\omega_A, \omega_B}$ and $|VV\rangle_{\omega_A, \omega_B}$ appear at the output of the cascade structure. Moreover, $\phi_{\mathcal{L},HV}(\omega_A, \omega_B)$ and $\phi_{\mathcal{L},VH}(\omega_A, \omega_B)$ [the prefactors of $|HV\rangle_{\omega_A, \omega_B}$ and $|VH\rangle_{\omega_A, \omega_B}$ in **Eq. (5)**] now contain contributions from both nonlinear segments, which eventually leads to interference between the biphoton amplitudes from the two nonlinear segments.

**Biphoton spectrum**

In this section, we assume there is no polarisation transformation in the middle section [*i.e.* $\hat{R}_n = e^{ik_n^{(0)}L_0}\begin{pmatrix} 1 & 0 \\ 0 & 1 \end{pmatrix}$ with $n=A, B$], and only focus on the spectrum of the biphotons generated from the cascade structure in that limit. For this case, if we assume the nonlinear media have $\Lambda(\omega_A, \omega_B) \ll 1$ [see **Eq. (6)**], the quantum state of the biphotons generated from the cascade structure becomes:

$$|\psi_{\text{CAS}}\rangle = \sum_{\mathcal{L}}\alpha_{\mathcal{L}}\int d\omega_P d\omega_A d\omega_B\, \mathcal{A}_0^{type\,II} f_{\mathcal{L}}(\omega_P)L_1 \text{sinc}\left(\frac{\Delta k_{VHV}^{(1)}L_1}{2}\right)\delta \times \\
\left(1 + U_{4P}e^{i(\Delta k^{(0)}L_0 + \Delta k_{VHV}^{(1)}L_1)}\right)\{|HV\rangle_{\omega_A, \omega_B} + |VH\rangle_{\omega_A, \omega_B}\}. \quad (6)$$

Here $U_{4P}$ is the fourth element of the transformation matrix $\widehat{U}_P$, which we write as $|U_{4P}|e^{i\phi_P}$. We now study the relative emission spectrum of the biphotons by expanding the total biphoton brightness $B_{\text{tot}} = \langle\psi_{\text{CAS}}|\psi_{\text{CAS}}\rangle$. As the expression for $B_{\text{tot}}$ involves the statistical phases of different pump packets, we first average $B_{\text{tot}}$ over the ensemble of sequences of the pump packets according to

$$\langle B_{\text{tot}}\rangle_{avg} = \int d\omega_P d\omega_A d\omega_B \langle\sum_{\mathcal{L}}|\alpha_{\mathcal{L}}f_{\mathcal{L}}(\omega_P)|^2\rangle_{avg}\left|\mathcal{A}_0^{type\,II}L_1\text{sinc}\left(\frac{\Delta k_{VHV}^{(1)}L_1}{2}\right)\right|^2\delta \times \\
\left|1 + U_{4P}e^{i(\Delta k^{(0)}L_0 + \Delta k_{VHV}^{(1)}L_1)}\right|^2. \quad (7)$$

We now replace the ensemble average $\langle\sum_{\mathcal{L}}|\alpha_{\mathcal{L}}f_{\mathcal{L}}(\omega_P)|^2\rangle_{avg}$ with $|\alpha|^2|f(\omega_P)|^2$, where $|\alpha|^2$ is the number of photons in the entire pump packets (see **supplement, section 8**); $|f(\omega_P)|^2$ is now the spectral lineshape of the pump, which is assumed to be Lorentzian[49]. Note that the integral of the form $\int d\omega_P e^{i(\Delta k^{(0)}L_0 + \Delta k_{VHV}^{(1)}L_1)}|f(\omega_P)|^2$ that appears in **Eq. (7)** is related to the first-order coherence of the pump, $g^{(1)}(\tau)$[see ref. 49]. Given that, we can re-write **Eq. (7)** as:

$$\langle B_{\text{tot}}\rangle_{avg} = \int d\omega_P d\omega_A d\omega_B\left|\mathcal{A}_0^{type\,II}\alpha f(\omega_P)L_1\text{sinc}\left(\frac{\Delta k_{VHV}^{(1)}L_1}{2}\right)\right|^2\delta \times \\
\left\{1 + |U_{4P}|^2 + 2|U_{4P}|e^{\frac{-|\Delta\tau_0 + \Delta\tau_1|}{\tau_c}}\cos(\Delta k^{(0)}L_0 + \Delta k_{VHV}^{(1)}L_1 + \Phi_P)\right\}, \quad (8)$$

where the factor $e^{\frac{-|\Delta\tau_0+\Delta\tau_1|}{\tau_C}}$ appears in **Eq. (8)** as a result of first-order coherence function of a pump field with coherence time of $\tau_c$ (see **supplement, section 8**); $\Delta\tau_0$ ($\Delta\tau_1$) is the group delay difference between pump and biphotons in the middle section (first nonlinear medium), which can be expressed as:

$$\Delta\tau_{0/1}=\tau_{0/1,P}-\frac{1}{2}\left(\tau_{0/1,A}+\tau_{0/1,B}\right)=\left(\left.\frac{dk^{(0/1)}}{d\omega}\right|_{\overline{\omega}_p}-\left.\frac{dk^{(0/1)}}{d\omega}\right|_{\frac{\overline{\omega}_p}{2}}\right)L_{0/1}, \quad (9)$$

where $\left.\frac{dk^{(0/1)}}{d\omega}\right|_{\omega'}$ is the first-order dispersion of the middle section/PPSF at frequency $\omega'$. The factor $e^{\frac{-|\Delta\tau_0+\Delta\tau_1|}{\tau_C}}$ in **Eq. (8)** determines to what extent the biphoton amplitudes from two nonlinear segments interfere. Note that the integrand in **Eq. (8)** corresponds to the biphoton spectrum (or joint spectral intensity). In the following subsections, we study the biphoton spectrum under two different pump coherence conditions.

*Biphoton spectrum: Coherent cascading*
When $|\Delta\tau_0+\Delta\tau_1| \leq \tau_C$, the pump field remains coherent throughout both nonlinear segments; we call this mode of operation "coherent cascading". Here the biphoton amplitudes from the two different nonlinear segments interfere with each other, resulting in fringes in the biphoton spectrum (see **supplement, section 8**). Adding more nonlinear segments (**Fig. 2a**) results in more interference terms, which gives us greater flexibility in shaping the biphoton spectrum. As an example, we could generate biphotons with discrete frequency modes (in the form of a frequency comb[20,22]) by cascading three PPSFs whose spectra are initially *continuous* (**Fig. 2b,c**). Note that the spacing between the frequency modes in **Fig. 2c** can be controlled by tailoring the dispersion of the middle section, without resorting to any spectral filtering or modification of the nonlinear media. It is also worth mentioning that since we are utilizing type-II SPDC and using PPSFs as our nonlinear media [$\Lambda(\omega_A,\omega_B) \ll 1$, **Eq. (3)**], biphotons generated from the cascade structure (**Fig. 2c**) are also entangled in the polarisation DOF as well[26,51].

When $|\Delta\tau_1| \leq \tau_C$ and the middle section has no dispersion (equivalent to $L_0=0$), the coherent cascade of two nonlinear segments becomes equivalent to a longer biphoton source with the total nonlinear interaction length of $L_{NL}=L_1+L_2=2L_1$. In this case, the brightness of the biphotons generated in the cascade structure increases by a factor of $2^{3/2}$ (scaling with $L_{NL}^{3/2}$, see ref. 48) with respect to the individual nonlinear segment, while the emission bandwidth [now determined by $\mathrm{sinc}^2\left(\Delta k_{VHV}^{(1)}L_1\right)$] is reduced by a factor of $2^{1/2}$ (scaling with $L_{NL}^{-1/2}$, see ref. 48) with respect to each of individual nonlinear segment. Note that both of the scaling factors mentioned here generally applies for degenerate SPDC processes in which the signal and idler have the same polarisation. However, as the group birefringence of PPSF is negligible[51] over the bandwidth of the downconverted photons (see **supplement, section 8.2**), the scaling factors mentioned above also apply for the type-II SPDC phase-matching in the case of PPSF. **Figure 3** shows the brightness (3a) and the emission bandwidth (3b) of $N$ identical PPSFs that are coherently cascaded. As can be seen in **Fig. 3b**, coherent cascade of multiple nonlinear segments (equivalent to increasing the length of the nonlinear medium) reduces the emission bandwidth of the biphotons, which is particularly undesirable for *broadband* biphoton sources; however, we will show in the following that this issue can be overcome through incoherent cascade of multiple nonlinear segments.

*Biphoton spectrum: Incoherent cascading*
When $|\Delta\tau_1| \leq \tau_C \ll |\Delta\tau_0|$, which we refer to as "incoherent cascading", the factor $e^{\frac{-|\Delta\tau_0+\Delta\tau_1|}{\tau_C}}$ in **Eq. (8)** vanishes and the biphoton brightness simplifies to:

$$\langle B_{tot}\rangle_{avg}=\int d\omega_P d\omega_A d\omega_B \left|\mathcal{A}_0^{type\,II}\alpha f(\omega_P)L_1 \mathrm{sinc}\left(\frac{\Delta k_{VHV}^{(1)}L_1}{2}\right)\right|^2 \delta \quad (10)$$
$$\times\{1+|U_{4P}|^2\}.$$

With such a low coherence pump, the biphoton amplitudes from the two nonlinear segments will not interfere, and the brightness of the biphotons becomes twice that of an individual nonlinear segment as $|U_{4P}| \to 1$. More generally, the brightness increases linearly with respect to the total nonlinear interaction length in the cascade structure. On the other hand, the emission bandwidth of the biphotons [determined by $\mathrm{sinc}^2\left(\frac{\Delta k_{VHV}^{(1)}L_1}{2}\right)$] remains the same as the emission bandwidth of an individual segment. As the number of identical cascaded nonlinear segments $N$ increases, the emission bandwidth remains constant, while the brightness increases linearly (**Fig. 3**). This suggests that with incoherent cascading, we can arbitrarily increase the brightness of the biphoton source without sacrificing the emission bandwidth of the biphotons which is in contrast to coherent cascading, where increasing the total nonlinear interaction length $L_{NL}$ was accompanied by a reduction in the emission bandwidth.

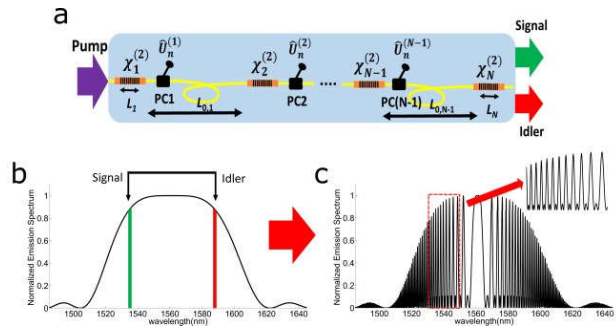

**Figure 2** (**a**) A generalized cascade structure consisting of $N$ nonlinear segments; $\hat{U}_n^{(1)}$, $\hat{U}_n^{(2)}$, and $\hat{U}_n^{(N-1)}$ are the polarisation transformation matrices of the PCs in the middle sections. The lengths of the nonlinear media and middle sections are denoted by $L_i$ and $L_{0,i}$, respectively. The emission spectrum of the biphotons generated from (**b**) a 20 cm PPSF, and (**c**) a cascade structure consisting of three identical 20 cm-long PPSFs connected with two 6 m-long SMF28™; The subset shows discretization of the frequency modes.

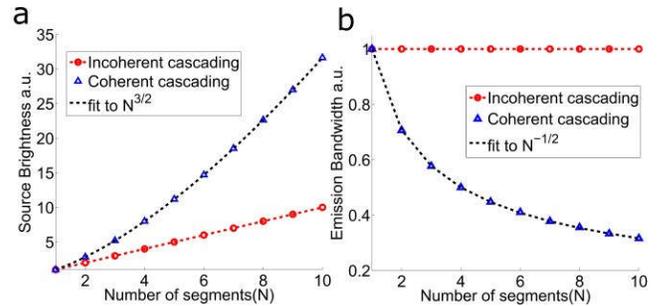

**Figure 3** Trade-off between the brightness and the emission bandwidth of the biphotons generated from the cascade structure with coherent and incoherent pumping of $N$ identical PPSFs. (**a**) Brightness scales linearly with respect to $N$ (or $L_{NL}$) for incoherent cascading, while it scales by factor of $N^{3/2}$ (or $L_{NL}^{3/2}$) for coherent cascading. (**b**) The emission bandwidth is independent of $N$ (or $L_{NL}$) for incoherent cascading, while it decreases with a factor of $N^{-1/2}$ (or $L_{NL}^{-1/2}$) for coherent cascading.

**Biphoton polarisation state**
In the following subsections, we study the effect of cascading on the degree of polarisation entanglement (quantified by concurrence[54]) and the polarisation state of the biphotons generated from the cascade structure. Although our approach can be applied to all SPDC phase-matching processes, in the interest of brevity, we discuss only the cases where type-II SPDC process occurs in both nonlinear media. As in previous cases, we account for the collective effects of the pump coherence, dispersion of the linear and nonlinear media, and the polarisation transformation applied in the middle section.

We model the unitary transformation of the middle section for signal and idler by a general unitary matrix of the form:

$$\widehat{U}_{A(B)}(\theta, \phi_1, \phi_2) = \begin{pmatrix} e^{i\phi_1}\cos\theta & -e^{i\phi_2}\sin\theta \\ e^{-i\phi_2}\sin\theta & e^{-i\phi_1}\cos\theta \end{pmatrix}, \quad (11)$$

where $\theta$ is the angle of the polarisation rotation; $\phi_1$ and $\phi_2$ are the phase parameters, which define an arbitrary polarisation transformation; note that, $\phi_1$ and $\phi_2$ physically correspond to the birefringence introduced by the optical elements in the middle section, such as the PC. The collective transformation matrix of the middle section then becomes $\widehat{R}_n = e^{ik_{0n}L_0}\widehat{U}_n(\theta, \phi_1, \phi_2)$. Given the polarisation transformation, we can use **Eq. (5)** and form the density matrix $\hat{\rho} = \langle B_{tot}\rangle_{avg}^{-1}|\psi_{CAS}\rangle\langle\psi_{CAS}|$ (with $\text{Tr}(\hat{\rho}) = 1$) to characterize the polarisation state of the biphotons generated from the cascade structure.

*Degree of polarisation entanglement*
In order to study the effect of cascading on the degree of polarisation entanglement, we limit ourselves to transformation of the form $\widehat{U}_n(\theta = \phi_1 = \phi_2 = 0)$. The ensemble-averaged density matrix of the biphoton state in polarisation bases ($|HH\rangle_{\omega_A,\omega_B}$, $|HV\rangle_{\omega_A,\omega_B}$, $|VH\rangle_{\omega_A,\omega_B}$, $|VV\rangle_{\omega_A,\omega_B}$) can be written as:

$$\langle \hat{\rho}_{\theta=\phi_1=\phi_2=0}^{typeII}\rangle_{avg} = \langle B_{tot}\rangle_{avg}^{-1}\int d\omega_P d\omega_A d\omega_B \left|\mathcal{A}_0^{typeII}\alpha f(\omega_P)L_1\text{sinc}\left(\frac{\Delta k_{VHV}^{(1)}L_1}{2}\right)\right|^2 \delta \times$$

$$\left\{1 + |U_{4P}|^2 + 2|U_{4P}|e^{\frac{-|\Delta\tau_0+\Delta\tau_1|}{\tau_C}}\cos(\Delta k^{(0)}L_0 + \Delta k_{VHV}^{(1)}L_1 + \Phi_P)\right\}\begin{pmatrix} 0 & 0 & 0 & 0 \\ 0 & 1 & e^{-i2\Lambda} & 0 \\ 0 & e^{i2\Lambda} & 1 & 0 \\ 0 & 0 & 0 & 0 \end{pmatrix}.$$
(12)

When $\int d\omega_A d\omega_B\, e^{i2\Lambda} = 0$, the density matrix for both pump coherence conditions has zero concurrence[54]. This is due to the walk-off between the two biphoton polarisation states $|HV\rangle_{\omega_A,\omega_B}$ and $|VH\rangle_{\omega_A,\omega_B}$, which is introduced by the nonlinear segments. The walk-off in the cascade structure can usually be compensated by placing a birefringent element in the path of biphotons. However, it is more desirable to use nonlinear media with $\Lambda \ll 1$ (such as poled-fibres[51]), especially when dealing with complex configurations[55] consisting of multiple cascaded nonlinear segments. The use of such nonlinear media ($\Lambda \ll 1$) in the cascade structure also allows us to, for example, preserve polarisation entanglement (if present) and simultaneously perform spectral shaping, similar to what we mentioned in previous sections.

*Shaping the polarisation state of the biphotons*
Now we study the role of polarisation transformation in shaping the polarisation state of the biphotons generated from the cascade structure. We consider a polarisation rotation of the form $\widehat{U}_n(\theta = \frac{\pi}{4}, \phi_1 = \phi_2 = 0)$, for which the density matrix of the biphoton state becomes:

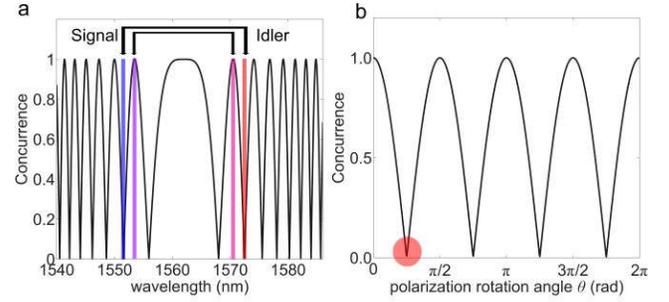

**Figure 4** (**a**) Concurrence as a function of signal and idler wavelengths for coherent cascading of two 30-cm-long PPSFs connected via a 3-m-long SMF28™ (used as the middle section). The polarisation transformation is set to $\widehat{U}_n(\theta = \frac{\pi}{4}, \phi_1 = \phi_2 = 0)$. For certain signal (idler) wavelengths, such as those denoted by blue (red) strips closer to the degeneracy point, the concurrence is 1, while for the adjacent strips, the concurrence is 0. (**b**) Concurrence as a function of the angle of the polarisation rotation in the middle section ($\theta$) for incoherent cascading; the designated red circle corresponds to $\widehat{U}_n(\theta = \frac{\pi}{4}, \phi_1 = \phi_2 = 0)$.

$$\langle \hat{\rho}_{\theta=\frac{\pi}{4},\phi_1=\phi_2=0}^{typeII}\rangle_{avg} = \langle B_{tot}\rangle_{avg}^{-1}\int d\omega_P d\omega_A d\omega_B \left|\mathcal{A}_0^{typeII}\alpha f(\omega_P)L_1\text{sinc}\left(\frac{\Delta k_{VHV}^{(1)}L_1}{2}\right)\right|^2 \delta \times$$

$$\begin{pmatrix} 1 & -\rho_1^* & -\rho_1^* & -1 \\ -\rho_1 & 1 & 1 & \rho_2 \\ -\rho_1 & 1 & 1 & \rho_2 \\ -1 & \rho_2^* & \rho_2^* & 1 \end{pmatrix},$$
(13)

where

$$\rho_1 = |U_{4P}|e^{\frac{-|\Delta\tau_0+\Delta\tau_1|}{\tau_C}}e^{i(\Delta k^{(0)}L_0+\Delta k_{VHV}^{(1)}L_1+\Phi_P-\Gamma_B)},$$

$$\rho_2 = |U_{4P}|e^{\frac{-|\Delta\tau_0+\Delta\tau_1|}{\tau_C}}e^{i(\Delta k^{(0)}L_0+\Delta k_{VHV}^{(1)}L_1+\Phi_P-\Gamma_A)}.$$
(14)

Note that we have assumed the nonlinear segments satisfy $\Lambda \ll 1$.

For coherent cascading ($|\Delta\tau_0 + \Delta\tau_1| \ll \tau_C$), the matrix in **Eq. (13)** is wavelength-dependent due to $\rho_1$ and $\rho_2$ elements [see **Eq. (14)**]. In fact, it can be shown that for a small wavelength range of the signal and idler photons, the concurrence varies between 0 and 1 (see **Fig. 4a**). Note that the signal and idler photons are still spectrally correlated, while the degree of polarisation entanglement varies with respect to signal and idler wavelengths. The effect shown in **Fig. 4a** is a direct consequence of *simultaneously* manipulating the dispersion and birefringence of the middle section.

For incoherent cascading ($|\Delta\tau_1| \leq \tau_C \ll |\Delta\tau_0|$), on the other hand, $\rho_1, \rho_2 \to 0$ and the density matrix has now zero concurrence for the entire signal and idler wavelength range. Here the variation of the concurrence is a result of polarisation rotation, and mixing of the two maximally polarisation-entangled biphoton states. In fact, it can be shown that by varying $\theta$ in $\widehat{U}_n(\theta, \phi_1 = \phi_2 = 0)$, we can vary the value of the concurrence between 0 and 1, obtaining a biphoton state with arbitrary degree of polarisation entanglement (see **Fig. 4b**). Note that here we only considered type-II SPDC for both pump coherence conditions; however, in practice, two biphoton sources with differing SPDC phase-matchings can be combined within our cascade structure to generate an *arbitrary* biphoton polarisation state.

## Experiment

A tunable cw tunable 780-nm external-cavity diode-laser (ECDL, Toptica DL PRO) with a coherence time of $\tau_C \approx 3$ μs (coherence length of $L_C \approx 1$ km) is used as a pump for coherent cascading. For incoherent cascading, we either decrease the time-averaged pump coherence by modifying the external cavity or separately pump the two PPSFs while the biphotons still travel in a common path (see **Methods** section). The pump power is adjusted for a pair generation rate $\leq 10^6$ pairs.s$^{-1}$, for which the probability of multi-pair generation is so small that it can be ignored. To demonstrate biphoton shaping in the spectral and polarisation domains, we add 5m of SMF28$^{TM}$ alongside an inline polarisation controller (PC2 in the inset of **Fig. 5**) to manipulate the dispersion and birefringence of the middle section.

For our proof-of-principle demonstration, three types of measurements are performed on each individual PPSF sample, as well as the cascade structure as a whole (**Fig. 5**): (1) Measurement of the biphoton spectrum to observe the spectral interference, and to obtain the emission bandwidth of the biphotons; (2) a coincidence measurement to quantify the biphoton brightness; and (3) quantum state tomography[56] (QST) to study the polarisation state of the biphotons under various polarisation transformations. The detection apparatus consists of two single photon detectors (SPDs, IDQ ID220), and a time-interval analyzer (TIA, Hydraharp 400). The biphoton spectrum is measured with an in-house fibre spectrometer[57] (**Fig. 5b**). The spectral resolution of our spectrometer is approximately 0.75 nm (100 GHz), limited primarily by the time jitter of our detectors.

For our setup, we choose two similar PPSFs by using the approach mentioned in the **Methods** section. The type-0 and type-II SPDC emission spectra of the two PPSFs are shown in **Fig. 6**. We observe the spectral overlap (~80 nm) is large around the degeneracy point, which allows us to obtain interference [predicted in **Eq. (8)**] in the spectrum of the biphotons generated from the cascade structure.

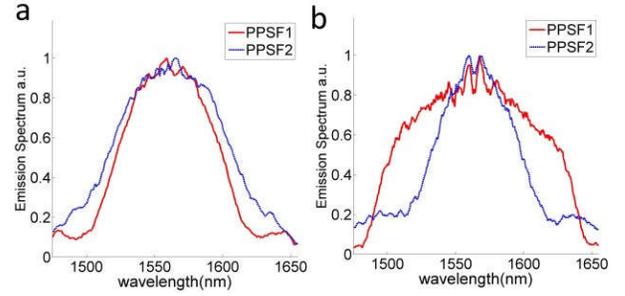

**Figure 6** Emission spectrum of each PPSF for (**a**) type-0 and (**b**) type-II SPDC; the emission spectra of the two segments are similar and overlap over a large bandwidth of ~80 nm.

### Biphoton spectral properties: Coherent cascading

We cascade the two PPSFs and perform type-0 and type-II SPDC, depending on the pump wavelength used (**Fig. 7**). The emission spectra of the biphotons at the output of the cascade structure are shown in **Fig. 7a,b**. Note that we have not yet applied any polarisation rotation in the middle section $[\widehat{U}_n(\theta = 0, \phi_1, \phi_2)$, see **Eq. (11)**]. The spectral interference fringes appear in the biphoton spectrum due to the dispersion of a 5-m-long SMF28$^{TM}$ in the middle section, which connects the two fiber-pigtailed PPSFs. The less-than-unity fringe visibility is mainly due to the spectral and brightness discrepancy between the two PPSFs. In our case, the fringe visibility of type-0 SPDC (~86%) is almost similar to what is observed for type-II SPDC (~81%) around the degeneracy wavelength, where the emission spectra of the two PPSFs are well matched (see **Fig. 6**). However, as we go further from the degeneracy wavelength, the discrepancy in the spectra of the two PPSFs increases (especially for type-II), and the fringe visibility of type-0 and type-II drops to ~80% and ~50%, respectively.

To highlight the effect of the pump coherence on the biphoton interference, we now decrease the time-averaged coherence of the pump (see **Methods**) and measure the spectrum of the biphotons at the output of the cascade structure for type-0 SPDC process. We use type-0 since the emission spectra of the two PPSFs are similar, so that the initial assumption of identical emission spectra of the two nonlinear sources holds true. The fringe visibility disappears in **Fig. 7a** (red trace), and the biphoton spectrum is now just an incoherent sum of the two individual PPSF spectra; this is in a good agreement with our simulation result shown in **Fig. 7c**.

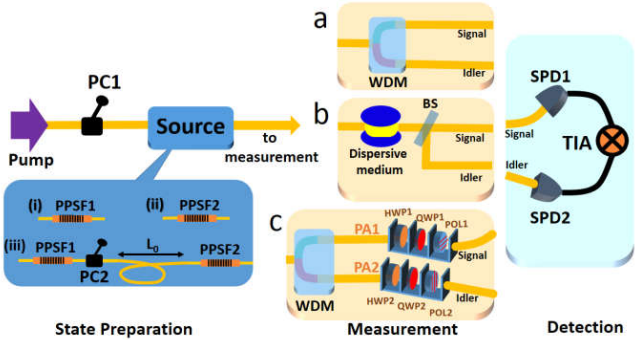

**Figure 5** Experimental setup. The source under test, illustrated in the inset, is pumped by a tunable cw diode laser. (**a**) A standard L-C band wavelength-division multiplexer (WDM) is used to separate signal (shorter wavelength) and idler (longer wavelength) photons into two different fibres for coincidence measurement. (**b**) For spectral measurements, a dispersive medium (20 km Corning SMF28$^{TM}$), and a beam splitter (BS) are used as a fibre spectrometer that extracts the biphoton spectrum. The overall detection time jitter is ~200 ps, based on which we select our coincidence time window to be 256 ps for spectral measurements. The nominal dispersion-length product of the 20 km fibre spool is 340 ps.nm$^{-1}$, which gives a spectral resolution of 0.75 nm in our measurement. (**c**) For the QST experiment, two sets of HP 8169A polarisation analyzers (PAs) are used, each of which includes a quarter waveplate (QWP), a half waveplate (HWP), and a polarizer (POL).

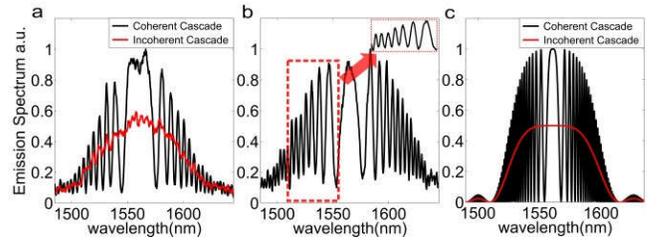

**Figure 7** (**a**) The emission spectrum of the biphotons generated from the cascade structure for type-0 SPDC (pumped at 780.05 nm) when $|\Delta\tau_1 + \Delta\tau_0| \ll \tau_C$ (coherent cascade), and $|\Delta\tau_1| \leq \tau_C \ll |\Delta\tau_0|$ (incoherent cascade, red trace). (**b**) The emission spectrum of the biphotons generated from the cascade structure for type-II SPDC, when coherently pumped at 782.05 nm; the inset shows the effect of the dispersion in the middle section on discretization of the frequency components. (**c**) Simulation result of the biphoton spectrum for type-0 SPDC cascade; the cascade structure consists of two identical PPSFs, each of length of 25 cm (= $L_1 = L_2$) connected by an SMF28$^{TM}$ patchcord of length $L_0 = 5$ m.

**Biphoton spectral properties: Incoherent cascading**

In this section, we study the brightness and emission spectrum of the biphotons generated from the cascade structure under incoherent pumping. To quantify the biphoton brightness [$\langle B_{tot}\rangle_{avg}$ in **Eq. (10)**], we measure the equivalent quantity, the coincidence rate of the biphotons (**see supplement, section 8**), for each individual sample as well as the cascade structure. We chose type-0 SPDC (pumped at 780.05 nm) since the emission spectra of the two PPSFs largely overlap (see **Fig. 6a**), allowing us to observe the variation in the emission bandwidth of the biphotons.

We first pump each PPSF and measure the coincidence rates with respect to the pump power. Taking into account the effect of loss for the pump and the signal (idler) fields, the expected coincidence rate for the cascade structure becomes:

$$R_{\exp} = \eta_{A(B),2}^2 R_{\text{PPSF1}} + \eta_{P,1} R_{\text{PPSF2}}, \quad (15)$$

where $\eta_{A(B),2}$ is the transmission of the signal (idler) field from the output of the PPSF1 to the output of the PPSF2, and $\eta_{P,1}$ is the transmission of the pump field from the input of PPSF1 to the input of PPSF2; $R_{\text{PPSF1}}$ and $R_{\text{PPSF2}}$ are the coincidence rate of the first and the second PPSF, respectively. We then use separate pumping technique (see **Methods** section) to ensure incoherent cascading and measure the coincidence rate of the biphotons generated from the cascade structure and compare it with $R_{\exp}$ in **Eq. (15)**. Note that the polarisation transformation in the middle section is set to $\widehat{U}_n(\theta=0,\phi_1,\phi_2)$ [see **Eq. (11)**] during our measurements. The result in **Fig. 8a** shows that for incoherent cascade, the brightness increases additively, and therefore scales linearly with the total nonlinear interaction length.

We then measure the emission spectrum of the biphotons generated from the cascade structure. As can be seen in **Fig. 8b**, the emission spectrum is the arithmetic mean of the two individual PPSF's spectra due to almost equal contribution of the two PPSFs at the output of the cascade structure. The result in **Fig. 8b** also shows no bandwidth reduction, which indicates that the emission bandwidth in incoherent cascading becomes independent of total nonlinear interaction length inside the cascade structure.

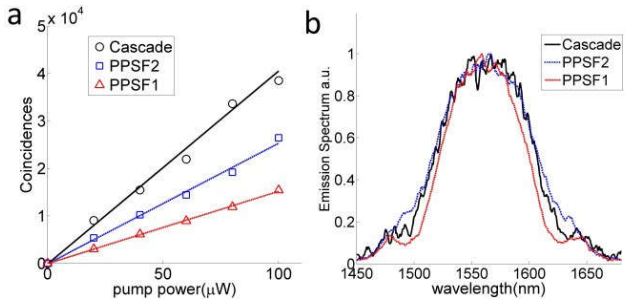

**Figure 8** (**a**) Coincidence rates are plotted as a function of pump power for type-0 SPDC. Symbols are measured data, while the solid and dashed lines are linear fits to the data points. For PPSF1, the displayed data points are the measured coincidence rates corrected by $\eta_{A(B),2}^2$. For PPSF2, the displayed data points are the measured coincidence rates corrected by $\eta_{P,1}$. The error bars are so small that they cannot be shown in the figure. (**b**) Type-0 SPDC spectra of the individual segment, and the cascaded. Due to nearly equal contribution of each PPSF sample in the output, the biphoton spectrum at the output of the cascade structure is the average of the two PPSF spectra, and shows no bandwidth reduction.

**Biphoton polarisation state: Incoherent cascading**

We now study the degree of polarisation entanglement and the polarisation state of the biphotons generated from the cascade structure by considering two specific transformations: (1) $\widehat{U}_n(\theta=0,\phi_1,\phi_2)$, and (2) $\widehat{U}_n(\theta=\frac{\pi}{4},\phi_1,\phi_2)$. We first characterize the polarisation state of the biphotons generated from each PPSF when pumped for type-II SPDC at 782.05 nm. Results in **Fig. 9a,b** show that both PPSFs generate biphoton states with a high concurrence, and high fidelity to triplet state $|\Psi^+\rangle$ (see also **Table 1**).

*Degree of polarisation entanglement*

The setup for cascading is similar to that of **Fig. 5**, except we separately pump the two PPSFs (see **Methods**). This method of pumping helps us to precisely control the pairwise contributions of each PPSF segment in the final quantum state and at the same time enables us to shape the polarisation state of the biphotons. We now change the settings of PC2 (see **Fig. 5**) so that there would be no polarisation rotation in the middle section [$\widehat{U}_n(\theta=0,\phi_1,\phi_2)$, see **Eq. (11)**] and then measure the biphoton state again. It can be seen from **Fig. 9c** that the measured density matrix corresponds to a highly polarisation-entangled state. Note that due to the negligible value of $\varLambda$ for the PPSFs, no walk-off is introduced between $|HV\rangle_{\omega_A,\omega_B}$ and $|VH\rangle_{\omega_A,\omega_B}$, and the degree of the polarisation entanglement remains unchanged after cascading.

*Shaping the polarisation state of the biphotons*

By applying a polarisation rotation of $\theta=\frac{\pi}{4}$ in the middle section $[\widehat{U}_n(\theta=\phi_1=\frac{\pi}{4},\phi_2=0)]$, the density matrix of the output state changes into the one shown in **Fig. 9d**. The concurrence drops to $\sim 0.1$ despite both PPSF segments individually generating high-concurrence polarisation-entangled biphotons. This value of the concurrence is consistent with the one predicted in **Fig. 4b** and suggests that, for $0 \leq \theta \leq \frac{\pi}{4}$, we can arbitrary tune the concurrence between 0 and 1. Note that here we have only considered type-II SPDC with two polarisation transformations; however, by applying various transformations $\widehat{U}_n(\theta,\phi_1,\phi_2)$ with PC2 in the middle section and utilizing different SPDC phase-matchings, one can generate any biphoton polarisation states.

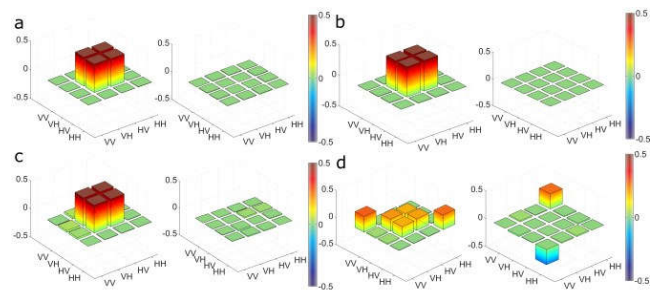

**Figure 9** Real and imaginary part of the output density matrix of (**a**) PPSF1 (**b**) PPSF2 and cascade structure corresponding to the polarisation transformation (**c**) $\widehat{U}_n(\theta=\phi_1=\phi_2=0)$ (**d**), and $\widehat{U}_n(\theta=\phi_1=\frac{\pi}{4},\phi_2=0)$ in the middle section. Note that the relative contributions of the two samples at the output are set to be similar (48% from PPSF1 and 52% from PPSF2) by adjusting the pump power for each one of them.

**Table 1** Fidelity, concurrence, and purity [Tr($\hat{\rho}^2$)] of the biphoton state measured in polarisation basis

| Output states | Fidelity to $|\Psi^+\rangle$ | Concurrence | Purity |
|---|---|---|---|
| PPSF1 | $98.5 \pm 0.2\%$ | $0.98 \pm 0.004$ | 0.988 |
| PPSF2 | $99.2 \pm 1.4\%$ | $0.99 \pm 0.02$ | 0.985 |
| Cascading under $\hat{U}_n(\theta = \phi_1 = \phi_2 = 0)$ | $99.2 \pm 1.8\%$ | $0.985 \pm 0.003$ | 0.986 |
| Cascading under $\hat{U}_n(\theta = \phi_1 = \frac{\pi}{4}, \phi_2 = 0)$ | $55 \pm 0.7\%$ | $0.1 \pm 0.01$ | 0.482 |

## Discussion

We have shown here that cascading biphoton sources in a *common-path* configuration can be used as a versatile tool to simultaneously tailor the frequency and polarisation DOFs of entangled photons. In this strategy, the pump coherence plays a major role in obtaining various biphoton states. With a long-coherence pump, the entire cascade structure can be considered as one unified source, capable of generating biphotons with tunable spectral properties[58-59]; in fact, one can obtain various biphoton spectra (**Fig. 7**) simply by engineering the dispersion of the linear medium in the middle section. For example, by cascading counter-propagating path-entangled biphoton sources[60], one could obtain biphoton frequency combs (similar to the works reported earlier[1,22]) of constant spacing whose free spectral range can be tuned by manipulating the dispersion of the middle section; note that, this can be done without any dispersion modification of the nonlinear medium.

Since our spectral and polarisation shaping techniques share the same configuration, we can simultaneously control biphotons in both DOFs. For example, by coherently pumping the cascade structure and manipulating the dispersion and birefringence property of the middle section, we can generate biphotons whose degree of polarisation entanglement is frequency-dependent (**Fig. 4a**); this new effect, arising from the interplay between coherence and entanglement, directly links the entanglement existing in the polarisation DOF to the frequency DOF of biphotons.

With incoherent pumping, the effects arising from biphoton-biphoton amplitude interference disappear, and the final state will become an incoherent mixture of the individual states generated from each nonlinear segment (see **Fig. 8**, **9**). The immediate application would be the ability to increase the brightness of the biphoton sources (at the expense of greater noise) by increasing the total nonlinear interaction length without sacrificing the emission bandwidth of the generated biphotons (see **Fig. 3**, **8**). In addition, the incoherent cascade scheme allows us to generate arbitrary biphoton polarisation states[13-14], and also control the degree of polarisation entanglement of biphotons (see **Fig. 4b** and **Fig. 9d**). We remark that our configuration greatly simplifies the schemes previously used for generating arbitrary biphoton polarisation state[13-14] and removes the requirement for phase stabilization and pump coherence due to its common-path configuration.

It is worth mentioning that using linear and nonlinear materials with negligible group birefringence (such as poled-fibres[51]) and small dispersion is of great importance in the cascade strategy as no walk-off between different biphoton polarisation states is essentially introduced. This feature allows us to preserve polarisation entanglement (if present), or generate arbitrary polarisation states[13-14] without the need for complex walk-off compensation schemes. This can also be beneficial for the configurations recently proposed for generating multi-photon entanglement[55], which often involve complex scheme of multiple cascaded nonlinear media. In addition, nonlinear media with small dispersion generate *broadband* biphotons and with the simple spectral shaping technique presented here, they can serve as versatile quantum sources for various quantum information processing applications[1-2].

The technique presented in this work can also be generalized to all other waveguide-based photon-pair sources, including those in integrated photonics devices, and one can use the effect of biphoton-biphoton amplitude interference to tune the properties of entangled photons, not only in the frequency and polarisation DOFs, but also in other DOFs such as path and orbital angular momentum as well. From this perspective, the cascade strategy can be invaluable for generating a host of entangled-photon states that could be useful for quantum information processing, quantum sensing, and the study of the foundations of quantum mechanics.

## Methods

### Choosing PPSF samples
The SHG spectrum of several PPSF samples are examined with a cw tunable laser (Agilent 8164 A) in the 1550-1565 nm wavelength range[61]. Depending on the input polarization of the fundamental lightwave, the type-0 or type-II phase-matching can be observed[61]. Two PPSFs whose SHG peaks and SPDC spectra are well-matched are then selected as nonlinear segments for our cascade structure.

### Fiber spectrometer
The biphotons generated in the cascade structure are sent to a dispersive medium (20 km of Corning SMF-28), which maps their wavelengths onto the arrival time at the single photon detectors[57]. After time-tagged detection with single photon detectors, the spectrum of the biphotons can be recovered by translating the time delays into wavelength[46,57]. The minimum resolution of our spectrometer depends primarily on the timing jitter of the single photon detectors[46]. As the overall detection time-jitter is ~200 ps, we chose our coincidence time window to be 256 ps for these measurements. Based on this coincidence window and the nominal dispersion-length product of the 20 km fibre spool, which is 340 ps.nm$^{-1}$, we can obtain spectral resolution of ~ 0.75 nm with our spectrometer.

### Incoherent pumping schemes
Depending upon suitability, one of two following methods is used to achieve incoherent pumping: (1) We decrease the effective coherence time of the pump laser by periodically modulating the cavity length of ECDL (the 780 nm laser) so that its time-averaged linewidth (as measured by a Fabry-Perot spectrometer) increases, effectively reducing the pump temporal coherence; (2) We pump the two PPSFs separately while the biphotons still travel in a common path. Since the two pump fields reaching the PPSF segments travel in two different and unstabilized fiber paths, no coherence is preserved between the two fields, guaranteeing incoherent cascading.

# Supplementary Materials

# Contents

In the first three sections, we lay out the foundation needed to derive an expression for electric fields in different regions of the the cascade structure. Then we use these fields for a Hamiltonian treatment of our scheme (**section 4-5**); afterwards, we derive an expression for the quantum state of the biphotons generated from the cascade structure (**section 6**). In the remaining sections(**section 7-9**), we study the biphoton state under the various pump coherence conditions and polarization transformations mentioned in the paper.

# 1 Field operators in the cascade structure

The displacement field vector associated with channel $n$ (*i.e.* $P$ for the pump, $A$ for the signal, and $B$ for the idler) can be written as:

$$\vec{D}_n(\vec{r}) = \sum_S \int dk_n \sqrt{\frac{\hbar \omega_{nk_n}}{4\pi}} a_{nSk_n} d_n(x,y) \vec{D}_{nSk_n}(z) + h.c.. \tag{1}$$

where $a_{nSk_n}$ is lowering operator associated with the field $n$ with polarization state of $S$ ($S=H,V$)[1]. We assume the mode shape $d_n(x,y)$ is independent of either polarization or $z$ and does not vary over the bandwidth of signal and idler. The vector $\vec{D}_{nSk}(z)$ in **Equation 1** has a negligible z component and corresponds to the Jones vector $\overline{D}_{nSk_n}(z)$ in $H$-$V$ basis. We choose to express $\overline{D}_{nSk_n}(z)$ in terms of the polarization of the field entering or leaving the cascade structure [1].

For the pump ($n=P$), we derive the Jones vector for different regions[$\overline{D}_{PSk_P}(z)$] based on the polarization vector of the pump field incident on the structure at $z=0$ (see **Fig.1a** in the paper). We call the Jones vector derived by this convention asymptotic-in vector (with superscript *in*), and denote it by $\overline{D}_{PSk_P}^{in}(z) = \begin{pmatrix} \overline{D}_{PSk_P}^{X-in}(z) \\ \overline{D}_{PSk_P}^{Y-in}(z) \end{pmatrix}$. For example, a vertically-polarized pump at $z=0$ can be expressed as $\overline{D}_{PVk_P}^{in}(z=0) = \begin{pmatrix} 0 \\ 1 \end{pmatrix}$. On the other hand, for the signal (idler), we derive $\overline{D}_{nSk_n}(z)$ based on the polarization vector of the field leaving the structure at $z=Z_f$(see **Fig.1a** in the paper). We call the Jones vector derived by this convention asymptotic-out vector (with superscript *out*), and for signal (idler), denote it by $\overline{D}_{ASk_A}^{out}(z) = \begin{pmatrix} \overline{D}_{ASk_A}^{X-out}(z) \\ \overline{D}_{ASk_A}^{Y-out}(z) \end{pmatrix} [\overline{D}_{BSk_B}^{out}(z) = \begin{pmatrix} \overline{D}_{BSk_B}^{X-out}(z) \\ \overline{D}_{BSk_B}^{Y-out}(z) \end{pmatrix}]$. For example, horizontally-polarized signal field at $z=Z_f$ can be expressed as $\overline{D}_{AHk_A}^{out}(z=Z_f) = \begin{pmatrix} 1 \\ 0 \end{pmatrix}$. Note that these fields are solutions to the Maxwell's equations. With the Jones vector $\overline{D}_{nSk_n}^{in/out}(z) = \begin{pmatrix} \overline{D}_{nSk_n}^{X-in/out}(z) \\ \overline{D}_{nSk_n}^{Y-in/out}(z) \end{pmatrix}$ for all regions in hand, we can write $\vec{D}_{nSk_n}(z)$ (in **Equation 1**) as:

$$\vec{D}_{nSk_n}^{in/out}(z) = \hat{x}\left[\overline{D}_{nSk_n}^{X-in/out}(z)\right] + \hat{y}\left[\overline{D}_{nSk_n}^{Y-in/out}(z)\right], \tag{2}$$

where $\hat{x}$ ($\hat{y}$) are the unit vectors in Cartesian coordinate system.

# 2 Quantum model of quasi-monochromatic field

We treat the pump as a quasi-monochromatic field, which is a succession of coherent packets[2] (**Figure 1a**). Within each packet $\mathcal{L}$ the electric field is a sinusoidal function with a wavenumber of $\overline{k}_p$(and angular frequency of $\overline{\omega}_p$), which is the same for all packets **(Figure 1a)**. We introduce a generalized pump operator $A_{\mathcal{L},S}$ for each coherent packet as:

$$A_{\mathcal{L},S} = \int dk_P f^*_\mathcal{L}(k_P) a_{PSk_P}, \qquad (3)$$

where $\mathcal{L}$ and $S$ indicate the packet number and packet polarization, respectively; $f_\mathcal{L}(k_P)$ contains the spectral behavior of packet $\mathcal{L}$. More specifically, $f_\mathcal{L}(k_P)$ is the Fourier transform of the spatial function $g_\mathcal{L}(z)$ of each packet (**see Figure 1b**); $g_\mathcal{L}(z)$ can be considered a rectangular function defined by $\sqrt{\frac{1}{2\pi\Delta_\mathcal{L}}} e^{i\Theta_\mathcal{L}} e^{i\bar{k}_p z} rect(\frac{z-z_\mathcal{L}}{\Delta_\mathcal{L}})$[2, 3], where $\Theta_\mathcal{L}$ is a phase that is statistically distributed for each packet[3]. Accordingly, we find

$$f_\mathcal{L}(k_P) = e^{i\Theta_\mathcal{L}} e^{-i(k_P-\bar{k}_p)z_\mathcal{L}} \sqrt{\frac{\Delta_\mathcal{L}}{2\pi}} sinc\left[(k_P-\bar{k}_p)\frac{\Delta_\mathcal{L}}{2}\right] \qquad (4)$$

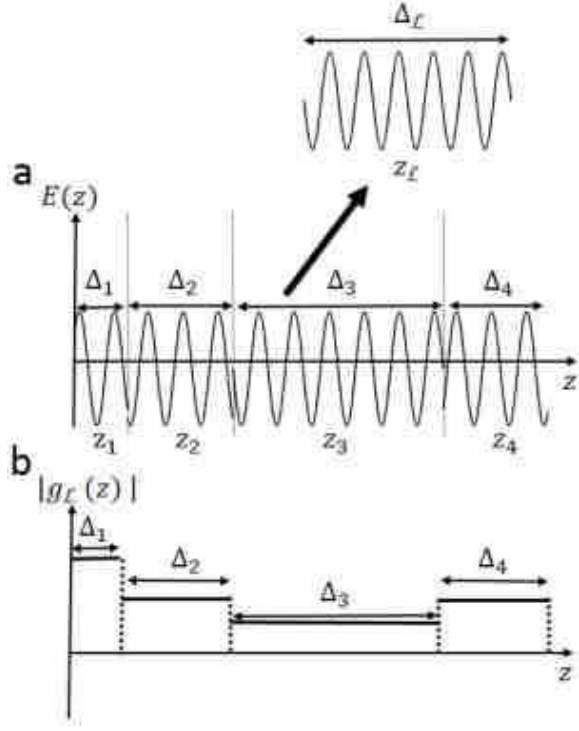

Figure 1: Model of the electric field of a laser with Lorentzian spectrum[2]; (a) The initial phase of the electrical field varies for each nonzero packet. The center of each packet is shown by $z_\mathcal{L}$ while the packet length is shown by $\Delta_\mathcal{L}$. (b) The function $|g_\mathcal{L}(z)|$ describes the spatial distribution of each coherent packet.

The operators $A_{\mathcal{L},S}$ and $A^\dagger_{\mathcal{L},S}$ (defined in **Equation 3**) must now satisfy the commutation relation $\left[A_{\mathcal{L},S}, A^\dagger_{\mathcal{L}',S}\right] = \delta_{\mathcal{L}\mathcal{L}'}$. Calculating the commutator, we obtain:

$$\left[A_{\mathcal{L},S}, A^\dagger_{\mathcal{L}',S}\right] = \int dk_P dk'_P f^*_\mathcal{L}(k_P) f_{\mathcal{L}'}(k'_P) \left[a_{PSk_P}, a^\dagger_{PSk'_P}\right] = \int dk_P f^*_\mathcal{L}(k_P) f_{\mathcal{L}'}(k_P) = \delta_{\mathcal{L}\mathcal{L}'}, \qquad (5)$$

Note that the coherent packets do not in general form a complete set of basis states, however, we assume that they can be extended to form a complete set[4]. This allows us to write:

$$a_{PSk_P} = \sum_\mathcal{L} f_\mathcal{L}(k_P) A_{\mathcal{L},S}. \qquad (6)$$

Now according to our model, the quantum state of each individual pump packet $\mathcal{L}$ can be described by coherent state $|\alpha_{\mathcal{L},S}\rangle = e^{\alpha_{\mathcal{L},S} A^\dagger_{\mathcal{L},S} - h.c.} |vac\rangle$, where $|\alpha_{\mathcal{L},S}|^2$ is the average photon number of that packet. Since the pump field is taken to be a succession of all the coherent packets and the field operators of the different packets commute, we can write down the initial quantum state of the pump field(at $t = t_0$ and $z = 0$) with $S$ polarization as

$$|\psi(t = t_0)\rangle = |\{\alpha_{\mathcal{L}.S}\}\rangle = e^{\sum_{\mathcal{L}} \alpha_{\mathcal{L},S} A^{\dagger}_{\mathcal{L},S} - h.c.} |vac\rangle. \tag{7}$$

For type II SPDC, which we consider in the paper, the polarization of the pump ($S$) incident on the structure will be always set to be $V$; however, as the pump propagates along the $z$-axis in the cascade structure, its polarization transforms. This effect is included in $\vec{D}_{PSk_P}(z)$ vector (**Equation 1**), which we calculate in the next section.

It is also worth mentioning that in order to have an accurate physical picture of any observable with this quantum model of the pump (including $\alpha_{\mathcal{L}}$ or $f_{\mathcal{L}}$), one needs to take ensemble average of that observable over all coherent pump packets. Since the pump is modeled with stationary and ergodic statistical properties, this ensemble averaging is equivalent to "long-time averages of the observable in a single experiment"[2, 3] .We postpone this step to **Section 8-9**, where we study the spectrum and density matrix of the biphoton generated in the cascade structure.

## 3 Derivation of Jones vectors

### 3.1 Transformation matrices:

The horizontal and vertical axes of the cascade structure are defined based on the principal axes of the nonlinear segments. For example, in the case of the periodically-poled silica fiber (PPSF) that we use in the paper, the horizontal axis is always defined as the poling direction of the PPSF [5]. The transformation matrix of each segment in **Figure 2** can be defined as $\hat{T}_n^{(m)}(z) = \begin{pmatrix} e^{ik_{nH}^{(m)} z} & 0 \\ 0 & e^{ik_{nV}^{(m)} z} \end{pmatrix}$, where $k_{nH}^{(m)}$ is a shorthand for $k_H^{(m)}(\omega_n)$, and superscript (m) indicates the segment in which the field is traveling; 0 for the middle section, 1 and 2 for the first and the second nonlinear segment, and no superscript for anywhere outside of the cascade structure. We model the optics of the middle section as an isotropic phase accumulation $[\hat{T}_n^{(0)}(z)]$. This is because in our experiment, we use a single mode fiber (SMF28) with negligible birefringence ($k_{nH}^{(0)} \approx k_{nV}^{(0)}$), which results in a phase accumulation matrix of the form $\hat{T}_n^{(0)}(z) = e^{ik_n^{(0)}z}\hat{I}$, where $\hat{I}$ is the identity matrix. The latter transformation is followed by a unitary polarization transformation $\hat{U}_n = \begin{pmatrix} U_{1n}(\omega) & U_{2n}(\omega) \\ U_{3n}(\omega) & U_{4n}(\omega) \end{pmatrix}$[6], which we apply through the polarization controller(PC) in the middle section(see **Figure 2**). Based on this, the collective transformation of the middle section becomes:

$$\hat{R}_n = \hat{T}_n^{(0)}(z)\hat{U}_n = e^{ik_n^{(0)}z} \begin{pmatrix} U_{1n}(\omega) & U_{2n}(\omega) \\ U_{3n}(\omega) & U_{4n}(\omega) \end{pmatrix}. \tag{8}$$

Note that due to weak wavelength-dependent birefringence of the middle section in the operating frequency range of the signal, idler, and pump, we have $\hat{U}_A = \hat{U}_B \neq \hat{U}_P$.

### 3.2 $\overline{D}_{nSk_n}^{in/out}(z)$ vectors:

In this section, we derive an expression for the vector $\overline{D}_{nSk_n}^{in/out}(z)$ mentioned in **section 1**. First, we write down $\overline{D}_{nSk_n}^{in}(z)$ in terms of field incident on the cascade structure at $z=0$ $[\overline{D}_{nSk_n}^{in}(z = 0)]$ and find

$$\overline{D}_{nSk_n}^{in}(z) = \begin{cases} \hat{T}_n^{(1)}(z)\overline{D}_{nSk_n}^{in}(z = 0) & 0 < z < L_1 \\ \hat{T}_n^{(2)}(z - L_0 - L_1)\hat{R}_n\hat{T}_n^{(1)}(L_1)\overline{D}_{nSk_n}^{in}(z = 0) & L_0 + L_1 < z < Z_f \end{cases}. \tag{9}$$

Note that we are only interested in the field expressions inside the two nonlinear segments, as they contribute to the nonlinear Hamiltonian of the cascade structure. At $z = Z_f$, $\overline{D}_{nSk_n}^{in}(z = Z_f)$ can be written as

$$\overline{D}_{nSk_n}^{in}(z = Z_f) = \left[\hat{T}_n^{(2)}(L_2)\hat{R}_n\hat{T}_n^{(1)}(L_1)\right] \overline{D}_{nSk_n}^{in}(z = 0). \tag{10}$$

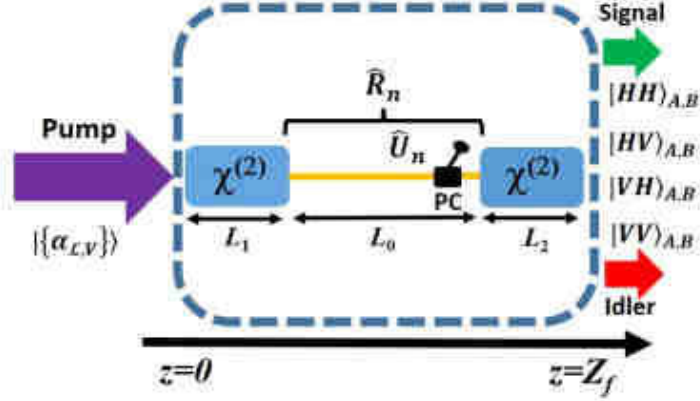

Figure 2: General cascade structure. The polarization state of the pump incident on the structure is $V$ shown by $|\{\alpha_{\mathcal{L},V}\}\rangle$. The output state of signal and idler at $z = Z_f$ has four different possible polarization states based on the total transformation in the middle section ($\hat{R}_n$).

In order to write down $\overline{D}_{nSk_n}^{out}(z)$ in terms of the outgoing field leaving the structure $[\overline{D}_{nSk_n}^{out}(z = Z_f)]$, we first use **Equation 10** and write

$$\overline{D}_{nSk_n}^{in}(z = 0) = \left[\hat{T}_n^{(2)}(L_2)\hat{R}_n\hat{T}_n^{(1)}(L_1)\right]^{-1}\overline{D}_{nSk_n}^{out}(z = Z_f). \tag{11}$$

Then, we substitute **Equation 11** back into **Equation 9** and get:

$$\overline{D}_{nSk_n}^{out}(z) = \begin{cases} \hat{T}_n^{(1)}(z)\left[\hat{T}_n^{(2)}(L_2)\hat{R}_n\hat{T}_n^{(1)}(L_1)\right]^{-1}\overline{D}_{nSk_n}^{out}(z = Z_f) & 0 < z < L_1 \\ \hat{T}_n^{(2)}(z - L_0 - L_1 - L_2)\overline{D}_{nSk_n}^{out}(z = Z_f) & L_0 + L_1 < z < Z_f \end{cases}. \tag{12}$$

Now, we derive an expression for $\overline{D}_{nSk_n}^{in/out}(z)$ when the polarization of the fields are set to either $H$ or $V$ at $z = 0$ (or $z = Z_f$). By setting $\overline{D}_{nSk_n}^{in}(z = 0)$ to $H$ or $V$, **Equation 9** becomes:

$$\overline{D}_{nHk_n}^{in}(z) = \begin{cases} \begin{pmatrix} e^{ik_{nH}^{(1)}z} \\ 0 \end{pmatrix} & 0 < z < L_1 \\ e^{ik_n^{(0)}L_0}e^{ik_{nH}^{(1)}L_1}\begin{pmatrix} U_{1n}e^{ik_{nH}^{(2)}(z-L_0-L_1)} \\ U_{3n}e^{ik_{nV}^{(2)}(z-L_0-L_1)} \end{pmatrix} & L_0 + L_1 < z < Z_f \end{cases}, \tag{13}$$

$$\overline{D}_{nVk_n}^{in}(z) = \begin{cases} \begin{pmatrix} 0 \\ e^{ik_{nV}^{(1)}z} \end{pmatrix} & 0 < z < L_1 \\ e^{ik_n^{(0)}L_0}e^{ik_{nV}^{(1)}L_1}\begin{pmatrix} U_{2n}e^{ik_{nH}^{(2)}(z-L_0-L_1)} \\ U_{4n}e^{ik_{nV}^{(2)}(z-L_0-L_1)} \end{pmatrix} & L_0 + L_1 < z < Z_f \end{cases}, \tag{14}$$

and by setting $\overline{D}_{nSk_n}^{out}(z = Z_f)$ to $H$ or $V$, **Equation 12** becomes:

$$\overline{D}_{nHk_n}^{out}(z) = \begin{cases} e^{-ik_n^{(0)}L_0}\begin{pmatrix} U_{4n}e^{-ik_{nH}^{(2)}L_2}e^{ik_{nH}^{(1)}(z-L_1)} \\ -U_{3n}e^{-ik_{nH}^{(2)}L_2}e^{ik_{nV}^{(1)}(z-L_1)} \end{pmatrix} & 0 < z < L_1 \\ \begin{pmatrix} e^{ik_{nH}^{(2)}(z-L_0-L_1-L_2)} \\ 0 \end{pmatrix} & L_0 + L_1 < z < Z_f \end{cases}, \tag{15}$$

$$\overline{D}_{nVk_n}^{out}(z) = \begin{cases} e^{-ik_n^{(0)}L_0} \begin{pmatrix} -U_{2n}e^{-ik_{nV}^{(2)}L_2}e^{ik_{nH}^{(1)}(z-L_1)} \\ U_{1n}e^{-ik_{nV}^{(2)}L_2}e^{ik_{nV}^{(1)}(z-L_1)} \end{pmatrix} & 0 < z < L_1 \\ \begin{pmatrix} 0 \\ e^{ik_{nV}^{(2)}(z-L_0-L_1-L_2)} \end{pmatrix} & L_0 + L_1 < z < Z_f \end{cases}. \tag{16}$$

## 4 Hamiltonian of the cascade structure

The Hamiltonian for the cascade structure is the sum of linear and the nonlinear parts

$$H_{tot} = H_L + H_{NL}, \tag{17}$$

where

$$H_L = \sum_S \left( \int dk_P \hbar\omega_{Pk_P} a_{PSk_P}^\dagger a_{PSk_P} + \int dk_A \hbar\omega_{Ak_A} a_{ASk_A}^\dagger a_{ASk_A} + \int dk_B \hbar\omega_{Bk_B} a_{BSk_B}^\dagger a_{BSk_B} \right). \tag{18}$$

and

$$H_{NL} = -\frac{1}{3\varepsilon_0} \int dr \eta^{ijk}(\vec{r}) D^i(\vec{r}) D^j(\vec{r}) D^k(\vec{r});$$

indices $i, j$, and $k$ refer to the Cartesian components of the fields; $\eta^{ijk}(\vec{r})$ can be written as[1]:

$$\eta^{ijk}(\vec{r}) = \frac{\chi_2^{ijk}(x,y)\Upsilon(z)}{\epsilon_0 n^2(x,y;\omega_P) n^2(x,y;\omega_A) n^2(x,y;\omega_B)}, \tag{19}$$

in which $\chi_2^{ijk}(x,y)$ is the second-order nonlinear tensor with no z-dependence, and $\Upsilon(z)$ is the spatial distribution of the nonlinearity along the z axis. For the cascade structure in **Figure 2**, $\Upsilon(z) = 1$ for $0 < z < L_1$ and $L_0 + L_1 < z < Z_f$, and $\Upsilon(z) = 0$ for all other regions. Now we rewrite $H_{NL}$ as

$$H_{NL} = -\frac{3!}{3\varepsilon_0} \sum_{S,S',S''} \int dk_P dk_A dk_B \mathfrak{s}^{S,S',S''}(k_P,k_A,k_B) a_{AS'k_A}^\dagger a_{BS''k_B}^\dagger a_{PSk_P} + h.c., \tag{20}$$

where $S$, $S'$, and $S''$ are the polarization of the pump, signal, and idler fields, respectively; $\mathfrak{s}^{S,S',S''}(k_p,k_A,k_B)$ can be described by

$$\mathfrak{s}^{S,S',S''}(k_P,k_A,k_B) = \sqrt{\frac{\hbar\omega_{Pk_P}\hbar\omega_{Ak_A}\hbar\omega_{Bk_B}}{(4\pi)^3}} \sum_{i,j,k} \int dr \eta^{ijk}(\vec{r}) d_P(x,y) d_A^*(x,y) d_B^*(x,y) D_{PSk_P}^{k-in}(z) \left( D_{AS'k_A}^{i-out}(z) D_{BS''k_B}^{j-out}(z) \right)^*. \tag{21}$$

Note that $\mathfrak{s}^{S,S',S''}(k_p,k_A,k_B)$ depends on the geometry and nonlinearity in the cascade structure [1]. Also, note that we have used asymptotic-in representation (**Equation 13-14**) for the pump field and asymptotic-out representation (**Equation 15-16**) for the signal and the idler fields. As an example, function $\mathfrak{s}^{V,H,V}(k_P,k_A,k_B)$ corresponds to the situation in which the polarization of the pump incident on the structure is $V$, while the polarization states of the signal and the idler fields generated at the output of the cascade structure are $H$ and $V$, respectively.

## 5 Calculating the nonlinear Hamiltonian

Given the asymptotic-in/out representation (**Equation 13-16**), we can calculate all $\mathfrak{s}^{S,S',S''}(k_P,k_A,k_B)$ and assemble $H_{NL}$ mentioned in **Equation 17**.

### 5.1 Calculating $\mathfrak{s}^{S,S',S''}$ function:

In calculating $\mathfrak{s}^{S,S',S''}$ we note:

1. The nonlinear media used are PPSFs[5] with four nonzero tensor components $\chi_2^{XXX}(x,y), \chi_2^{XYY}(x,y), \chi_2^{YXY}(x,y)$, and $\chi_2^{YYX}(x,y)$. The relation between these tensors is: $\frac{1}{3}\chi_2^{XXX}(x,y) = \chi_2^{XYY}(x,y) = \chi_2^{YXY}(x,y) = \chi_2^{YYX}(x,y)$ [5].

2. When pumping for a specific type of SPDC, only that type will get phase-matched and there would be no radiation from other types. This is due to different phase-matching wavelengths of different SPDC types. For example, for the PPSFs used in our experiments, the type-0 wavelength is at 780.05 nm while type-II wavelength is at 782.05 nm.

3. $k_{nS}^{(m)}$ is a shorthand for $k_S^{(m)}(\omega_n)$ and corresponds to wavenumber of field $n$ with polarization state $S$ inside medium $m$.

4. We use $\Delta k_{SS'S''}^{(m)} = k_{PS}^{(m)} - k_{AS'}^{(m)} - k_{BS''}^{(m)} - k_{QPM}$ to denote the wavenumber mismatch of each nonlinear segment $m$; $S$, $S'$ and $S''$ are polarization states of the pump, signal, and idler; $k_{QPM}$ is the quasi phase-matching wavenumber of PPSF; $k_{QPM} = 0$ for the middle section and we simply have $\Delta k^{(0)} = k_P^{(0)} - k_A^{(0)} - k_{B'}^{(0)}$; note, we removed the polarization indexes for the wavenumber in the middle section as the medium used is polarization independent.

5. For the terms that are slowly-varying in the angular frequency, we will replace $\omega_{Pk_P}$ with $\overline{\omega}_p$ (angular frequency of the cw pump) and $\omega_{Ak_A}$, $\omega_{Bk_B}$ with $\overline{\omega}_p/2$. Note that we use this approximation for all terms except phase factors and sinusoidal functions.

## 5.2 Type II SPDC cascade:

In type II SPDC, vertically-polarized pump photons are down-converted into two cross-polarized photons, one with horizontal and the other with the vertical polarization. In order to calculate $\mathfrak{s}^{S,S',S''}(k_P, k_A, k_B)$ for the cascade structure, we take the spatial integral (see **Equation 21**) over the entire structure from $z=0$ to $z=Z_f$. According to our assumptions, when the structure is pumped for type II SPDC, only the terms including $\Delta k_{VHV}^{(m)}$ and $\Delta k_{VVH}^{(m)}$ would survive and the remaining terms give no contribution when spatial integral is taken. As the polarization of the pump incident on the structure (at $z=0$) is vertical, we set $S=V$ in **Equation 21** and write $\mathfrak{s}^{VHH}(k_P, k_A, k_B)$ as

$$\mathfrak{s}^{VHH}(k_P, k_A, k_B) = \mathfrak{A}_0^{(typeII)} e^{-i(k_{BV}^{(2)} - k_{BH}^{(2)})L_1} \left\{ (-U_{4A}U_{3B})^* L_1 sinc(\Delta k_{VHV}^{(1)} L_1/2) + \right.$$
$$\left. e^{i(k_{AV}^{(1)} - k_{AH}^{(1)} + k_{BH}^{(1)} - k_{BV}^{(1)})\frac{L_1}{2}} (-U_{3A}U_{4B})^* L_1 sinc(\Delta k_{VVH}^{(1)} L_1/2) \right\} \quad (22)$$

where

$$\mathfrak{A}_0^{(typeII)} = -\frac{3!}{6\varepsilon_0} \int dxdy \sqrt{\left(\frac{\hbar\overline{\omega}_p}{4\pi}\right)^3} \frac{\chi_2^{YXY}(x,y) d_P(x,y) d_A^*(x,y) d_B^*(x,y)}{\epsilon_0 n^2(x,y;\overline{\omega}_p) n^2(x,y;\frac{\overline{\omega}_p}{2}) n^2(x,y;\frac{\overline{\omega}_p}{2})} \times$$
$$e^{i(k_{AH}^{(1)} + k_{BV}^{(1)})L_1} e^{i(k_{AH}^{(2)} + k_{BH}^{(2)})L_2} e^{i(k_A^{(0)} + k_B^{(0)})L_0} e^{i\frac{\Delta k_{VHV}^{(1)} L_1}{2}}.$$

Note that we assumed in the beginning that $\chi_2^{YXY}(x,y) = \chi_2^{YYX}(x,y)$[5]; we also used the slowly-varying assumption and replaced $\omega_p$ and $\omega_A(\omega_B)$ with $\overline{\omega}_p$ and $\frac{\overline{\omega}_p}{2}$, respectively; according to this, we find

$$\mathfrak{s}^{VHV}(k_P, k_A, k_B) = \mathfrak{A}_0^{(typeII)} \left\{ (U_{4A}U_{1B})^* L_1 sinc(\Delta k_{VHV}^{(1)} L_1/2) + e^{i(k_{AV}^{(1)} - k_{AH}^{(1)} + k_{BH}^{(1)} - k_{BV}^{(1)})\frac{L_1}{2}} (U_{3A}U_{2B})^* L_1 sinc(\Delta k_{VVH}^{(1)} L_1/2) \right.$$
$$\left. + U_{4P} e^{i\Delta k^{(0)} L_0} e^{i\Delta k_{VHV}^{(1)} L_1} L_2 sinc(\Delta k_{VVH}^{(2)} L_2/2) \right\}, \quad (23)$$

and

$$\mathfrak{s}^{VVH}(k_P, k_A, k_B) = \mathfrak{A}_0^{(typeII)} e^{i(k_{AV}^{(1)} - k_{AH}^{(1)} + k_{BH}^{(1)} - k_{BV}^{(1)})L_2} \left\{ (U_{2A}U_{3B})^* L_1 sinc(\Delta k_{VHV}^{(1)} L_1/2) + \right.$$
$$\left. e^{i(k_{AV}^{(1)} - k_{AH}^{(1)} + k_{BH}^{(1)} - k_{BV}^{(1)})\frac{L_1}{2}} (U_{1A}U_{4B})^* L_1 sinc(\Delta k_{VVH}^{(1)} L_1/2) + U_{4P} e^{i(k_{AV}^{(1)} - k_{AH}^{(1)} + k_{BH}^{(1)} - k_{BV}^{(1)})\frac{L_2}{2}} e^{i\Delta k^{(0)} L_0} e^{i\Delta k_{VHV}^{(1)} L_1} L_2 sinc(\Delta k_{VVH}^{(2)} L_2/2) \right\},$$
$$(2$$

and finally

$$\mathfrak{s}^{VVV}(k_P, k_A, k_B) = \mathfrak{A}_0^{(type\ II)} e^{i(k_{AV}^{(2)} - k_{AH}^{(2)})L_2} \left\{ (-U_{2A}U_{1B})^* L_1 sinc(\Delta k_{VHV}^{(1)} L_1/2) + \right.$$
$$\left. e^{i(k_{AV}^{(1)} - k_{AH}^{(1)} + k_{BH}^{(1)} - k_{BV}^{(1)})\frac{L_1}{2}} (-U_{1A}U_{2B})^* L_1 sinc(\Delta k_{VVH}^{(1)} L_1/2) \right\}. \quad (25)$$

Now we assume the two nonlinear media have a similar dispersion properties *i.e.* $k_{nS}^{(1)} = k_{nS}^{(2)}$. We can also assume $L_1 = L_2$, and $sinc(\Delta k_{VHV}^{(1)} L_1/2) \approx sinc(\Delta k_{VVH}^{(1)} L_1/2)$[7] over the phase-matching wavelength range, which results in $H_{NL}$ of the form:

$$H_{NL} = \int dk_P dk_A dk_B \mathfrak{A}_0^{(type\ II)} L_1 sinc(\Delta k_{VHV}^{(1)} L_1/2) \times \left\{ -e^{-i\Gamma_B} \left[ (U_{4A}U_{3B})^* + e^{i\Lambda}(U_{3A}U_{4B})^* \right] a_{pVk_P} a_{AHk_A}^\dagger a_{BHk_B}^\dagger + \right.$$
$$\left[ (U_{4A}U_{1B})^* + e^{i\Lambda}(U_{3A}U_{2B})^* + U_{4P} e^{i\Delta k^{(0)} L_0} e^{i\Delta k_{VHV}^{(1)} L_1} \right] a_{pVk_P} a_{AHk_A}^\dagger a_{BVk_B}^\dagger +$$
$$e^{2i\Lambda} \left[ (U_{2A}U_{3B})^* + e^{i\Lambda}(U_{1A}U_{4B})^* + e^{-i\Lambda} U_{4P} e^{i\Delta k^{(0)} L_0} e^{i\Delta k_{VHV}^{(1)} L_1} \right] a_{pVk_P} a_{AVk_A}^\dagger a_{BHk_B}^\dagger +$$
$$\left. -e^{i\Gamma_A} \left[ (U_{2A}U_{1B})^* + e^{i\Lambda}(U_{1A}U_{2B})^* \right] a_{pVk_P} a_{AVk_A}^\dagger a_{BVk_B}^\dagger + h.c. \right\}. \quad (26)$$

where $\Lambda = (k_{AV}^{(1)} - k_{AH}^{(1)} + k_{BH}^{(1)} - k_{BV}^{(1)})\frac{L_1}{2}$, $\Gamma_B = (k_{BV}^{(1)} - k_{BH}^{(1)})L_1$, and $\Gamma_A = (k_{AV}^{(1)} - k_{AH}^{(1)})L_1$.

## 6 Quantum state of the cascade structure

Given the $H_{NL}$ from the previous section, we calculate the output quantum state of the cascade structure in the interaction picture. First, we write down the time-dependent Schrodinger equation

$$i\hbar \frac{d}{dt} |\psi_S(t)\rangle = H_{tot} |\psi_S(t)\rangle \quad (27)$$

where $H_{tot}$ is the total Hamiltonian mentioned in **Equation 17**; $|\psi_S(t)\rangle$ is the state vector in Schrodinger picture, which is also a solution to **Equation 27**. To work in interaction picture, we define $|\psi_I(t)\rangle = e^{iH_L(t-t_0)/\hbar} |\psi_S(t)\rangle$, where $|\psi_I(t)\rangle$ is the state vector in the interaction picture; note that for $t = t_0$, we have $|\psi_I(t_0)\rangle = |\psi_S(t_0)\rangle$. By substituting $|\psi_S(t)\rangle = e^{-iH_L(t-t_0)/\hbar} |\psi_I(t)\rangle$ in **Equation 27,** we find that the time-evolution of the state vector in the interaction picture ($|\psi_I(t)\rangle$) can be describe by a Schrodinger equation of the form

$$i\hbar \frac{d}{dt} |\psi_I(t)\rangle = H_{NL}(t) |\psi_I(t)\rangle \quad (28)$$

with $H_{NL}(t) = e^{iH_L(t-t_0)/\hbar} H_{NL} e^{-iH_L(t-t_0)/\hbar}$. By solving **Equation 28** up to the first-order of perturbation we get:

$$|\psi_I(t)\rangle = \left( \hat{I} + \frac{1}{i\hbar} \int_{t_0}^t dt' H_{NL}(t') + ... \right) |\psi_I(t_0)\rangle, \quad (29)$$

where $H_{NL}(t')$ is

$$H_{NL}(t') = \sum_{S,S',S''} \int dk_P dk_A dk_B \mathfrak{s}^{S,S',S''}(k_P, k_A, k_B) e^{-i(\omega_P - \omega_A - \omega_B)t'} a_{AS'k_A}^\dagger a_{BS''k_B}^\dagger a_{PSk_P} + h.c. .$$

In the limit of $t_0 = -\infty$ and $t = \infty$, the time integral in **Equation 29** gives energy conservation $\delta(\omega_P - \omega_A - \omega_B)$. Now we can write down the quantum state at the output of the cascade structure, $|\psi_{out}\rangle = |\psi_I(t)\rangle$, as:

$$|\psi_{out}\rangle \approx \left[ \hat{I} - i\frac{2\pi}{\hbar} \times \right.$$
$$\left. \left( \sum_{S,S',S''} \int dk_P dk_A dk_B \mathfrak{s}^{S,S',S''}(k_P, k_A, k_B) \delta(\omega_P - \omega_A - \omega_B) a_{AS'k_A}^\dagger a_{BS''k_B}^\dagger a_{PSk_P} + h.c. + ... \right) \right] |\psi_I(t_0)\rangle, \quad (30)$$

where $|\psi_I(t_0)\rangle = e^{\sum_\mathcal{M} \alpha_{\mathcal{M},\mathcal{H}} A^\dagger_{\mathcal{M},\mathcal{H}} - h.c.} |vac\rangle$ (see **Equation 7**). Replacing $a_{PSk_P}$ with $\sum_\mathcal{L} f_\mathcal{L}(k_P) A_{\mathcal{L},S}$ for the pump, the term $\sum_\mathcal{L} f_\mathcal{L}(k_P) A_{\mathcal{L},S} \times e^{\sum_\mathcal{M} \alpha_{\mathcal{M},\mathcal{H}} A^\dagger_{\mathcal{M},\mathcal{H}} - h.c.} |vac\rangle$ appears in **Equation 30**. Now using lemma $\left[A_{\mathcal{L},S}, F(A^\dagger_{\mathcal{M},\mathcal{H}})\right] = \delta_{\mathcal{L}\mathcal{M}} \delta_{S\mathcal{H}} \frac{\partial F(A^\dagger_{\mathcal{M},\mathcal{H}})}{\partial A^\dagger_{\mathcal{M},\mathcal{H}}}$, we get:

$$\sum_\mathcal{L} f_\mathcal{L}(k_P) A_{\mathcal{L},S} \times e^{\sum_\mathcal{M} \alpha_{\mathcal{M},\mathcal{H}} A^\dagger_{\mathcal{M},\mathcal{H}} - h.c.} |vac\rangle = \sum_\mathcal{L} \sum_\mathcal{M} \delta_{\mathcal{L}\mathcal{M}} \delta_{S\mathcal{H}} f_\mathcal{L}(k_P) \alpha_{\mathcal{M},\mathcal{H}} \times e^{\sum_\mathcal{M} \alpha_\mathcal{M} A^\dagger_{\mathcal{M},\mathcal{H}} - h.c.} |vac\rangle = \sum_\mathcal{L} \alpha_{\mathcal{L},S} f_\mathcal{L}(k_P) |\psi_I(t_0)\rangle. \tag{31}$$

Substituting back into **Equation 30** we obtain

$$|\psi_{out}\rangle \approx \left[\hat{I} - i\frac{2\pi}{\hbar} \times \right.$$
$$\left. \left(\sum_\mathcal{L} \sum_{S,S',S''} \alpha_{\mathcal{L},S} \int dk_P dk_A dk_B f_\mathcal{L}(k_P) \mathfrak{s}^{S,S',S''}(k_P, k_A, k_B) \delta(\omega_P - \omega_A - \omega_B) a^\dagger_{AS'k_A} a^\dagger_{BS''k_B} + h.c. + ...\right)\right] |\psi_I(t_0)\rangle. \tag{32}$$

Since the polarization of the pump field incident on the structure (at $z=0$) is always set to vertical for type-II SPDC, we replace $S=V$ and remove $\sum_S$ from **Equation 32**. From now on for convenience in our notations, we drop the pump polarization index from $\alpha_{\mathcal{L},S}$. Since we always post-select the photon pairs and filter out the pump field, the quantum states of the *photon pairs* becomes:

$$\left|\psi^{type\,II}_{out}\right\rangle \approx |vac\rangle - i\frac{2\pi}{\hbar} \left(\sum_\mathcal{L} \alpha_\mathcal{L} \sum_{S',S''} \int dk_P dk_A dk_B f_\mathcal{L}(k_P) \mathfrak{s}^{V,S',S''}(k_P, k_A, k_B) \delta(\omega_P - \omega_A - \omega_B) a^\dagger_{AS'k_A} a^\dagger_{BS''k_B}\right) |vac\rangle. \tag{33}$$

By replacing $a^\dagger_{AS'k_A} a^\dagger_{BS''k_B} |vac\rangle$ with $\left|S'S''\right\rangle_{\omega_A,\omega_B}$ and neglecting the vacuum contribution, we get:

$$\left|\psi^{type\,II}_{out}\right\rangle = \sum_\mathcal{L} \alpha_\mathcal{L} \sum_{S',S''} \int dk_A dk_B \phi^{type\,II}_{\mathcal{L},S'S''}(k_A, k_B) \left|S'S''\right\rangle_{\omega_A,\omega_B}. \tag{34}$$

where $\left|S'S''\right\rangle_{\omega_A,\omega_B}$ is the biphoton state (signal and idler photons) in polarization state $S'(S'')$. The biphoton wave function $\phi_{\mathcal{L},S'S''}$ for type II SPDC can then be expressed as:

$$\phi^{type\,II}_{\mathcal{L},S'S''}(k_A, k_B) = -i\frac{2\pi}{\hbar} \sum_{S',S''} \int dk_P f_\mathcal{L}(k_P) \mathfrak{s}^{V,S',S''}(k_P, k_A, k_B) \delta(\omega_P - \omega_A - \omega_B). \tag{35}$$

# 7 Calculating $G^{(2)}(t_1, t_2)$

As we mentioned in the paper, we use coincidence detection to measure the brightness and emission spectrum of the biphotons generated in the cascade structure. Hence, the goal of the two following sections is to show how the coincidence is related to the total biphoton brightness $B_{tot} = \left\langle \psi^{type\,II}_{out} \middle| \psi^{type\,II}_{out} \right\rangle$ defined in the paper. We first calculate the second-order correlation function $G^{(2)}$ [8], which is proportional to the probability of the coincidence detection at the output of the cascade structure. The function $G^{(2)}$ we wish to calculate is of the form

$$G^{(2)}_{type\,II}(t_1, t_2) = \left\langle \Psi^{type\,II}_\mathcal{H} \middle| F^\dagger_A(\vec{r}_1, t_1) F^\dagger_B(\vec{r}_2, t_2) F_A(\vec{r}_1, t_1) F_B(\vec{r}_2, t_2) \middle| \Psi^{type\,II}_\mathcal{H} \right\rangle, \tag{36}$$

where $t_1$ and $t_2$ are the detection times of the signal and idler, respectively; $\left|\Psi_{\mathcal{H}}^{type\,II}\right\rangle = |\psi(t=t_0)\rangle$ (see **Equation 7**) is the initial quantum state of the pump incident on the structure in the Heisenberg picture; $F_A(\vec{r}_1, t_1)$ and $F_B(\vec{r}_2, t_2)$ are the field operators in the Heisenberg picture at the signal and idler detectors ($r_1$ and $r_2$) and can be expressed as

$$F_A(\vec{r}, t) = \sum_{S'} \int dk_A \sqrt{\frac{\hbar \overline{\omega}_p}{8\pi}} a_{AS'k_A}(t) d_A(x,y) e^{ik_A z}, \tag{37}$$

$$F_B(\vec{r}, t) = \sum_{S''} \int dk_B \sqrt{\frac{\hbar \overline{\omega}_p}{8\pi}} a_{BS''k_B}(t) d_B(x,y) e^{ik_B z}. \tag{38}$$

Note that in our experiment, the detectors do not resolve the mode shape $[\int dxdy d_A(x,y) = 1]$.

## 7.1 Field operators in Heisenberg picture:

Now we derive an expression for the field operators in **Equation 37-38** by calculating $a_{AS'k_A}(t)$ and $a_{BS''k_B}(t)$. Heisenberg equation of motion for the pump, signal, and idler can be written as:

$$i\hbar \frac{da_{nSk_n}}{dt} = \hbar \omega_{nk_n} a_{nSk_n} + [a_{nSk_n}, H_{NL}], \tag{39}$$

By neglecting pump depletion, we can write **Equation 39** for different polarization states of the signal and idler fields as:

$$i\hbar \frac{da_{AS'k_A}(t)}{dt} = \hbar \omega_{Ak_A} a_{AS'k_A}(t) + \sum_{S''} \int dk_P dk_B \mathfrak{s}^{V,S',S''}(k_P, k_A, k_B) a^\dagger_{BS''k_B}(t) a_{PVk_P}(t), \tag{40}$$

$$i\hbar \frac{da_{BS''k_B}(t)}{dt} = \hbar \omega_{Bk_B} a_{BS''k_B}(t) + \sum_{S'} \int dk_P dk_A \mathfrak{s}^{V,S',S''}(k_P, k_A, k_B) a^\dagger_{AS'k_A}(t) a_{PVk_P}(t). \tag{41}$$

Substituting $a_{PVk_P}(t) = a_{PVk_P} e^{-i\omega_{Pk_P}(t-t_0)}$ back into **Equation 40-41** and keeping the terms up to first order in perturbation theory, the solutions for the signal and idler fields become:

$$a_{AS'k_A}(t) = \left[a_{AS'k_A} - \frac{i}{\hbar}\sum_{S''}\int\int_{t_0}^t dt' dk_P dk_B \mathfrak{s}^{V,S',S''}(k_P, k_A, k_B) e^{i\Delta_{ABP}(t'-t_0)} a^\dagger_{BS''k_B} a_{PVk_P}\right] e^{-i\omega_{k_A}(t-t_0)} \tag{42}$$

$$a_{BS''k_B}(t) = \left[a_{BS''k_B} - \frac{i}{\hbar}\sum_{S'}\int\int_{t_0}^t dt' dk_P dk_A \mathfrak{s}^{V,S',S''}(k_P, k_A, k_B) e^{i\Delta_{ABP}(t'-t_0)} a^\dagger_{AS'k_A} a_{PVk_P}\right] e^{-i\omega_{k_B}(t-t_0)} \tag{43}$$

where $a_{PVk_P}$, $a^\dagger_{BS''k_B}$, and $a^\dagger_{AS'k_A}$ in the right hand side are Schrodinger operators, and $\Delta_{ABP} = \omega_{Ak_A} + \omega_{Bk_B} - \omega_{Pk_P}$. From now on, we denote $\mathfrak{s}^{V,S',S''}(k_P, k_A, k_B)$ by a short format $\mathfrak{s}^{V,S',S''}_{PAB}$. In order to include the pump packets, we substitute $a_{PVk_P}$ with $\sum_{\mathcal{L}} f_{\mathcal{L}}(k_P) A_{\mathcal{L},V}$ from **Equation 6** and define a new quantity $T^{V,S',S''}_{\mathcal{L}AB}(t)$ as

$$T^{V,S',S''}_{\mathcal{L}AB}(t) = -\frac{i}{\hbar}\int\int_{t_0}^t dt' dk_P \mathfrak{s}^{V,S',S''}_{PAB} e^{i\Delta_{ABP}(t'-t_0)} f_{\mathcal{L}}(k_P). \tag{44}$$

Now we can write **Equation 42-43** as:

$$a_{AS'k_A}(t) = \left[a_{AS'k_A} + \sum_{\mathcal{L}}\sum_{S''}\int dk_B T^{V,S',S''}_{\mathcal{L}AB}(t) a^\dagger_{BS''k_B} A_{\mathcal{L},V}\right] e^{-i\omega_{k_A}(t-t_0)} \tag{45}$$

$$a_{BS''k_B}(t) = \left[a_{BS''k_B} + \sum_{\mathcal{L}}\sum_{S'}\int dk_A T^{V,S',S''}_{\mathcal{L}AB}(t) a^\dagger_{AS'k_A} A_{\mathcal{L},V}\right] e^{-i\omega_{k_B}(t-t_0)}. \tag{46}$$

Table 1: Important parameters used in our notation.

| | symbol | Equivalent to |
|---|---|---|
| group delay of the pump in the nonlinear segment | $\tau_{1P}$ | $\left.\frac{dk^{(1)}}{d\omega}\right|_{\overline{\omega}_p} L_1$ |
| group delay of the signal/idler time in the nonlinear segment | $\tau_{1A/1B}$ | $\left.\frac{dk^{(1)}}{d\omega}\right|_{\frac{\overline{\omega}_p}{2}} L_1$ |
| group delay of the pump in the middle section | $\tau_{0P}$ | $\left.\frac{dk^{(0)}}{d\omega}\right|_{\overline{\omega}_p} L_0$ |
| group delay of the signal/idler time in the middle section | $\tau_{0A/0B}$ | $\left.\frac{dk^{(0)}}{d\omega}\right|_{\frac{\overline{\omega}_p}{2}} L_0$ |
| coherence time of the pump | $\tau_C$ | $\langle \tau_{\mathcal{L}} \rangle_{avg}$ |
| peak angular frequency of Lorentzian pump | $\overline{\omega}_p$ | |
| pump average power | $\mathfrak{p}$ | |

We now replace **Equation 45-46** into **Equation 37-38** and obtain

$$F_A(z_1, t_1) = \sqrt{\frac{\hbar\overline{\omega}_p}{8\pi}} \sum_{S'} \int dk_A \left[ a_{AS'k_A} + \sum_{\mathcal{L}} \sum_{S''} \int dk_B T_{\mathcal{L}AB}^{V,S',S''}(t) a_{BS''k_B}^\dagger A_{\mathcal{L},V} \right] e^{-i\omega_{k_A}(t_1-t_0)} e^{ik_A z_1}, \quad (47)$$

$$F_B(z_2, t_2) = \sqrt{\frac{\hbar\overline{\omega}_p}{8\pi}} \sum_{S''} \int dk_B \left[ a_{BS''k_B} + \sum_{\mathcal{L}} \sum_{S'} \int dk_A T_{\mathcal{L}AB}^{V,S',S''}(t) a_{AS'k_A}^\dagger A_{\mathcal{L},V} \right] e^{-i\omega_{k_B}(t_2-t_0)} e^{ik_B z_2}. \quad (48)$$

Note that we used slowly-varying criteria and replaced $\omega_{Ak_A}$ and $\omega_{Bk_B}$ with $\frac{\overline{\omega}_p}{2}$.

## 7.2 $G^{(2)}(t_1, t_2)$ function for type II SPDC cascade:

By substituting the field operators of **Equation 47-48** back into **Equation 36**, for type II SPDC we get

$$G_{typeII}^{(2)}(t_1, t_2) = \left(\frac{\hbar\overline{\omega}_p}{8\pi}\right)^2 \sum_{S',S''} \sum_{R',R''} \left[ \int dk_A dk_B dk_C dk_D e^{i(k_A-k_D)z_1} e^{i(k_B-k_C)z_2} e^{i(\omega_{Ak_A}-\omega_{Ak_D})t_1} e^{i(\omega_{Bk_B}-\omega_{Bk_C})t_2} \times \right.$$

$$\left[ \sum_{\mathcal{L},\mathcal{M}} \alpha_\mathcal{L} \alpha_\mathcal{M}^* T_{\mathcal{L}DC}^{V,S',S''}(t_1) T_{\mathcal{M}AB}^{V,R',R''}(t_1)^* + \right.$$

$$\left. \sum_{\mathcal{L},\mathcal{M},\mathcal{N},\mathcal{P}} \int dk_E dk_F \alpha_\mathcal{L} \alpha_\mathcal{M} \alpha_\mathcal{N}^* \alpha_\mathcal{P}^* \left( T_{\mathcal{P}FB}^{V,R',R''}(t_2) T_{\mathcal{N}AE}^{V,S',S''}(t_1) \right)^* \left( T_{\mathcal{M}FC}^{V,R',R''}(t_2) T_{\mathcal{L}DE}^{V,S',S''}(t_1) \right) \right]. \quad (49)$$

where dummy variables $A/D$ and $B/C$ correspond to the signal and idler fields, respectively. Note that the first term in the bracket corresponds to the actual coincidences, while the second terms corresponds to the accidental coincidences (or background noise).

# 8 Biphoton spectrum

## 8.1 General expression for biphoton brightness:

In practice, the total number of detected coincidences between signal and idler detectors can be calculated as a time integral of $G^{(2)}(t_1, t_2)$ over the entire measurement time $[\int dt_1 dt_2 G^{(2)}(t_1, t_2)]$. Since we always subtract the accidental coincidences, we disregard the second term of $G^{(2)}(t_1, t_2)$ in **Equation 49** and define a new quantity $\tilde{B}$ as

$$\tilde{B} = \int dt_1 dt_2 \int dk_A dk_B dk_C dk_D \sum_{S',S''} \sum_{R',R''} \left(\frac{\hbar \overline{\omega}_p}{8\pi}\right)^2 e^{i(k_A - k_D)z_1} e^{i(k_B - k_C)z_2} e^{i(\omega_{Ak_A} - \omega_{Ak_D})t_1} e^{i(\omega_{Bk_B} - \omega_{Bk_C})t_2} \times$$
$$\sum_{\mathcal{L},\mathcal{M}} \alpha_\mathcal{L} \alpha_\mathcal{M}^* T_{\mathcal{L}DC}^{V,S',S''}(t_1) T_{\mathcal{M}AB}^{V,R',R''}(t_1)^*. \quad (50)$$

We show that $\tilde{B}$ is proportional to the biphoton brightness $B_{tot}$ defined in the paper (**see Eqn. 7 in the paper**); this quantity $\tilde{B}$ also reveals the spectral brightness of the biphotons generated in the cascade structure. From now on to simplify our calculations, we change from the momentum space representation to the angular frequency representation whenever it is necessary. In order to do that, we replace $dk$ with $\frac{dk}{d\omega}|_{\omega'} d\omega$. For the signal/idler, we define $\beta_{1f} = \frac{dk}{d\omega}|_{\frac{\overline{\omega}_p}{2}}$, and for the pump frequency range we define $\beta_{1P} = \frac{dk}{d\omega}|_{\overline{\omega}_p}$. Now in the angular frequency representation, we can rewrite **Equation 50** as:

$$\tilde{B} = \sum_{\mathcal{L},\mathcal{M}} \alpha_\mathcal{L} \alpha_\mathcal{M}^* \sum_{S',S''} \sum_{R',R''} (2\pi \beta_{1f}^2)^2 \left(\frac{\hbar\overline{\omega}_p}{8\pi}\right)^2 \int d\omega_A d\omega_B T_{\mathcal{L}AB}^{V,S',S''}(t_1) T_{\mathcal{M}AB}^{V,R',R''}(t_1)^*. \quad (51)$$

By expanding $T_{\mathcal{L}DB}^{V,S',S''}(t_1)$ and $T_{\mathcal{M}AB}^{V,R',R''}(t_1)$ we have:

$$\tilde{B} = \left(\beta_{1P} \beta_{1f}^2\right)^2 \sum_{\mathcal{L},\mathcal{M}} \alpha_\mathcal{L} \alpha_\mathcal{M}^* \sum_{S',S''} \sum_{R',R''} \left(\frac{\overline{\omega}_p}{8\pi}\right)^2 \int \int_{t_0}^{t_1} \int_{t_0}^{t_1} d\omega_A d\omega_B dt' dt'' \int d\omega_P d\omega_{P'} \times$$
$$\mathfrak{s}_{PAB}^{V,S',S''} \left(\mathfrak{s}_{P'AB}^{V,R',R''}\right)^* e^{i\Delta_{ABP}(t'-t_0)} e^{-i\Delta_{ABP'}(t''-t_0)} f_\mathcal{L}(\omega_P) f_\mathcal{M}^*(\omega_{P'}). \quad (52)$$

Integrating over $t'$ and $t''$ we find

$$\tilde{B} = \left(\frac{2\pi\overline{\omega}_p \beta_{1P} \beta_{1f}^2}{4}\right)^2 \sum_{\mathcal{L},\mathcal{M}} \alpha_\mathcal{L} \alpha_\mathcal{M}^* \sum_{S',S''} \sum_{R',R''} \int d\omega_A d\omega_B \int d\omega_P d\omega_{P'} \mathfrak{s}_{PAB}^{V,S',S''} \left(\mathfrak{s}_{P'AB}^{V,R',R''}\right)^* \times$$
$$f_\mathcal{L}(\omega_P) f_\mathcal{M}^*(\omega_{P'}) \delta(\omega_B + \omega_A - \omega_P) \delta(\omega_B + \omega_A - \omega_{P'}), \quad (53)$$

which results in $\omega_P = \omega_{P'}$. Rewriting **Equation 53** we obtain

$$\tilde{B} = \gamma \sum_{S',S''} \sum_{R',R''} \sum_{\mathcal{L},\mathcal{M}} \int d\omega_A d\omega_B \alpha_\mathcal{L} \alpha_\mathcal{M}^* \int d\omega_P \mathfrak{s}_{PAB}^{V,S',S''} \left(\mathfrak{s}_{P'AB}^{V,R',R''}\right)^* \delta(\omega_P - \omega_A - \omega_B) f_\mathcal{L}(\omega_P) f_\mathcal{M}^*(\omega_P) \quad (54)$$

where $\gamma = \left(\frac{\pi \overline{\omega}_p \beta_{1P} \beta_{1f}^2}{2}\right)^2$. In the next section, we study the effect of the pump coherence on quantity $\tilde{B}$ in **Equation 54**.

## 8.2 Biphoton brightness in the absence of polarization transformation

To avoid complexity, we assume that there is no polarization rotation in the middle section $[\hat{U}_{A,B} = \begin{pmatrix} 1 & 0 \\ 0 & 1 \end{pmatrix}]$. Therefore, only $\mathfrak{s}_{PAB}^{V,H,V}$ and $\mathfrak{s}_{PAB}^{V,V,H}$ are nonzero and we can write ensemble average of **Equation 54** as

$$\left\langle \tilde{B} \right\rangle_{avg} = \gamma \int d\omega_A d\omega_B \left\langle \sum_{\mathcal{L},\mathcal{M}} \alpha_\mathcal{L} \alpha_\mathcal{M}^* \int d\omega_P f_\mathcal{L}(\omega_P) f_\mathcal{M}^*(\omega_P) \delta(\omega_P - \omega_A - \omega_B) \times \right.$$
$$\left. \left(\left|\mathfrak{s}_{PAB}^{V,H,V}\right|^2 + \left|\mathfrak{s}_{PAB}^{V,V,H}\right|^2 + 2Re\left\{\mathfrak{s}_{PAB}^{V,H,V} \left(\mathfrak{s}_{P'AB}^{V,V,H}\right)^*\right\}\right) \right\rangle_{avg} \quad (55)$$

where

$$\mathfrak{s}_{PAB}^{V,H,V} = \mathfrak{A}_0^{(typeII)} L_1 sinc(\Delta k_{VHV}^{(1)} L_1/2) \left[1 + U_{4P} e^{i\Delta k^{(0)} L_0} e^{i\Delta k_{VHV}^{(1)} L_1}\right], \tag{56}$$

$$\mathfrak{s}_{PAB}^{V,V,H} = \mathfrak{A}_0^{(typeII)} L_1 sinc(\Delta k_{VHV}^{(1)} L_1/2) e^{2i\Lambda} \left[e^{i\Lambda} + e^{-i\Lambda} U_{4P} e^{i\Delta k^{(0)} L_0} e^{i\Delta k_{VHV}^{(1)} L_1}\right]; \tag{57}$$

$U_{4P}$ is the fourth component of the polarization transformation matrix of the pump $|\hat{U}_P = \begin{pmatrix} U_{1P} & U_{2P} \\ U_{3P} & U_{4P} \end{pmatrix}$, see **section 3.1**] that can be written as $U_{4P} = |U_{4P}| e^{i\phi_P}$. Note that the definition of $f_\mathcal{L}(\omega_P)$ in **Equation 4** requires[3]

$$\left\langle \sum_{\mathcal{L},\mathcal{M}} \alpha_\mathcal{L} \alpha_\mathcal{M}^* f_\mathcal{L}(\omega_P) f_\mathcal{M}^*(\omega_P) \right\rangle_{avg} = \delta_{\mathcal{L},\mathcal{M}} \left\langle \sum_\mathcal{L} |\alpha_\mathcal{L} f_\mathcal{L}(\omega_P)|^2 \right\rangle_{avg}$$

Now we use stationary and ergodic statistical property of the pump[3] and find

$$\left\langle \sum_\mathcal{L} |\alpha_\mathcal{L}|^2 f_\mathcal{L}(\omega_P) \right\rangle_{avg} \approx |\alpha|^2 |f(\omega_P)|^2, \tag{58}$$

where $|f(\omega_P)|^2$ is the lineshape of the pump with Lorentzian spectrum, and $|\alpha|^2$ is the total average number of pump photons that can be written as

$$|\alpha|^2 = \frac{\mathfrak{p} T_{on}}{\hbar \bar{\omega}_P}, \tag{59}$$

where $\mathfrak{p}$ is the average pump power and $T_{on}$ is the pump-on time during the coincidence measurement. Since we are using PPSF, $\Lambda \ll 1$([7]) and we have $\mathfrak{s}_{PAB}^{V,H,V} = \mathfrak{s}_{PAB}^{V,V,H}$. This helps us to rewrite **Equation 55** as:

$$\left\langle \tilde{B} \right\rangle_{avg} = \gamma \int d\omega_A d\omega_B d\omega_P \left| \alpha \mathfrak{A}_0^{(type\,II)} L_1 sinc(\frac{\Delta k_{VHV}^{(1)} L_1}{2}) f(\omega_P) \left\{1 + U_{4P} e^{i\left(\Delta k^{(0)} L_0 + \Delta k_{VHV}^{(1)} L_1\right)}\right\} \right|^2 \delta(\omega_P - \omega_A - \omega_B). \tag{60}$$

In the following section, we show that the term $\int d\omega_P e^{i\left(\Delta k^{(0)} L_0 + \Delta k_{VHV}^{(1)} L_1\right)} |f(\omega_P)|^2$ in **Equation 60** contains the effect of pump coherence and modifies the spectrum of the biphotons (see the integrand in **Equation 60**) generated in the cascade structure.

### 8.2.1 The effect of pump coherence on spectral brightness of biphotons:

To show the effect of pump coherence, which is explicitly hidden in $\int d\omega_P e^{i\left(\Delta k^{(0)} L_0 + \Delta k_{VHV}^{(1)} L_1\right)} |f(\omega_P)|^2$ in **Equation 60**, we first Taylor expand $\Delta k^{(0)} L_0$ and $\Delta k_{VHV}^{(1)} L_1$ up to the first order about the degeneracy point of the pump ($\bar{\omega}_P$) and down-converted light ($\frac{\bar{\omega}_P}{2}$) and obtain:

$$\Delta k^{(0)}(\omega_P, \omega_A, \omega_B) L_0 = \left\{k_P^{(0)} - k_A^{(0)} + k_B^{(0)}\right\} L_0 =$$
$$\left\{k^{(0)}(\bar{\omega}_p) - k^{(0)}(\frac{\bar{\omega}_p}{2}) - k^{(0)}(\frac{\bar{\omega}_p}{2}) + \frac{dk^{(0)}}{d\omega}\bigg|_{\bar{\omega}_p} \Delta\omega_P - \frac{dk^{(0)}}{d\omega}\bigg|_{\frac{\bar{\omega}_p}{2}} (\Delta\omega_A + \Delta\omega_B) + ...\right\} L_0 \tag{61}$$

where $\Delta\omega_{A,B} = \omega_{A,B} - \frac{\bar{\omega}_p}{2}$, while $\Delta\omega_P = \omega_P - \bar{\omega}_p$ (for $\bar{\omega}_P$, see **table 1**). We again remind the reader that $k_n^{(m)}$ is a shorthand for $k^{(m)}(\omega_n)$. Due to energy conservation we have $\omega_P = \omega_A + \omega_B$, which leads to $\Delta\omega_A + \Delta\omega_B = \Delta\omega_P$. Putting this into **Equation 61** we get:

$$\Delta k^{(0)}(\omega_P,\omega_A,\omega_B)L_0 = \left\{ k^{(0)}(\bar{\omega}_p) - k^{(0)}(\frac{\bar{\omega}_p}{2}) - k^{(0)}(\frac{\bar{\omega}_p}{2}) + \left( \frac{dk^{(0)}}{d\omega}\bigg|_{\bar{\omega}_p} - \frac{dk^{(0)}}{d\omega}\bigg|_{\frac{\bar{\omega}_p}{2}} \right) \Delta\omega_P + ... \right\} L_0. \quad (62)$$

Similarly for $\Delta k^{(1)}_{VHV} L_1$ we get

$$\Delta k^{(1)}_{VHV}(\omega_P,\omega_A,\omega_B)L_1 = \left\{ k^{(1)}_{PV} - k^{(1)}_{AH} - k^{(1)}_{BV} - k_{QPM} \right\} L_1 =$$
$$\left\{ k^{(1)}_V(\bar{\omega}_p) - k^{(1)}_H(\frac{\bar{\omega}_p}{2}) - k^{(1)}_V(\frac{\bar{\omega}_p}{2}) - k_{QPM} + \left( \frac{dk^{(1)}_V}{d\omega}\bigg|_{\bar{\omega}_p} - \frac{dk^{(1)}_V}{d\omega}\bigg|_{\frac{\bar{\omega}_p}{2}} \right) \Delta\omega_P - \right.$$
$$\left. \left( \frac{dk^{(1)}_H}{d\omega}\bigg|_{\frac{\bar{\omega}_p}{2}} - \frac{dk^{(1)}_V}{d\omega}\bigg|_{\frac{\bar{\omega}_p}{2}} \right) \Delta\omega_A + ... \right\} L_1 \quad (63)$$

where $k^{(1)}_V(\bar{\omega}_p) - k^{(1)}_H(\frac{\bar{\omega}_p}{2}) - k^{(1)}_V(\frac{\bar{\omega}_p}{2}) - k_{QPM} = 0$ due to phase-matching constraint in PPSF. The first terms in **Equation 62-63** introduce a constant phase, however the second terms are the ones that affect the spectral interference. For simplicity, we can assume $\frac{dk^{(1)}_H}{d\omega}\bigg|_{\frac{\bar{\omega}_p}{2}} \approx \frac{dk^{(1)}_V}{d\omega}\bigg|_{\frac{\bar{\omega}_p}{2}}$, which results in canceling the last term in **Equation 63**; this happens to be a good approximation as PPSF has negligible group birefringence[7]. Now by using **Equation 62-63**, we rearrange integral $\int d\omega_P e^{i\left(\Delta k^{(0)} L_0 + \Delta k^{(1)}_{VHV} L_1\right)} |f(\omega_P)|^2$ into the standard format of the form $\int d\omega_P e^{i\omega_p \tau} |f(\omega_P)|^2$ that can be written as [3]

$$\int d\omega_P e^{i\omega_P \tau} |f(\omega_P)|^2 = g^{(1)}(\tau) \int d\omega_p |f(\omega_P)|^2, \quad (64)$$

where $g^{(1)}(\tau)$ is the first-order coherence function of the pump[2, 3], and $\tau$ is the time delay that has a units of time. For a pump field with Lorentzian lineshape centered around $\bar{\omega}_p$, $g^{(1)}(\tau) = e^{i\bar{\omega}_p \tau} e^{-\left|\frac{\tau}{\tau_C}\right|}$ where $\tau_C = \langle \tau_{\mathcal{L}} \rangle_{avg}$ is the coherence time of the pump field (see **table 1**). After some algebra and separating the terms of interest, we can rewrite $\int d\omega_P e^{i\left(\Delta k^{(0)} L_0 + \Delta k^{(1)}_{VHV} L_1\right)} |f(\omega_P)|^2$ as:

$$\int d\omega_P A(\bar{\omega}_p) e^{i\omega_P \left[\frac{dk^{(0)}}{d\omega}\big|_{\bar{\omega}_p} - \frac{dk^{(0)}}{d\omega}\big|_{\frac{\bar{\omega}_P}{2}}\right] L_0} e^{i\omega_P \left[\frac{dk^{(1)}_V}{d\omega}\big|_{\bar{\omega}_p} - \frac{dk^{(1)}_V}{d\omega}\big|_{\frac{\bar{\omega}_P}{2}}\right] L_1} |f(\omega_P)|^2, \quad (65)$$

where $A(\bar{\omega}_P)$ contains all the phase factors that are not function of $\omega_P$. Now with the help of **Equation 64**, we can write **Equation 65** as:

$$\int d\omega_P A(\bar{\omega}_p) e^{i\left(\Delta k^{(0)} L_0 + \Delta k^{(1)}_{VHV} L_1\right)} |f(\omega_P)|^2 = \int d\omega_P A(\bar{\omega}_p) |f(\omega_P)|^2 g^{(1)}(\tau) =$$
$$\int d\omega_P A(\bar{\omega}_P) |f(\omega_P)|^2 e^{-\left|\frac{\tau_{0P} - \frac{\tau_{0A}}{2} - \frac{\tau_{0B}}{2} + \tau_{1P} - \frac{\tau_{1A}}{2} - \frac{\tau_{1B}}{2}}{\tau_C}\right|} e^{i\bar{\omega}_p \left|\tau_{0P} - \frac{\tau_{0A}}{2} - \frac{\tau_{0B}}{2} + \tau_{1P} - \frac{\tau_{1A}}{2} - \frac{\tau_{1B}}{2}\right|}; \quad (66)$$

note that $\tau = \left|\tau_{0P} - \frac{\tau_{0A}}{2} - \frac{\tau_{0B}}{2} + \tau_{1P} - \frac{\tau_{1A}}{2} - \frac{\tau_{1B}}{2}\right|$, where $\tau_{0P} = \frac{dk^{(0)}}{d\omega}\bigg|_{\bar{\omega}_p} L_0$ and $\tau_{0A} = \tau_{0B} = \frac{dk^{(0)}}{d\omega}\bigg|_{\frac{\bar{\omega}_p}{2}} L_0$ while $\tau_{1P} = \frac{dk^{(1)}}{d\omega}\bigg|_{\bar{\omega}_p} L_1$ and $\tau_{1A} = \tau_{1B} = \frac{dk^{(1)}}{d\omega}\bigg|_{\frac{\bar{\omega}_p}{2}} L_1$. In fact, according to **Equation 66**, time delay $\tau$ is a function of *group delay* difference between the pump and biphotons (signal and idler) in the middle section as well as the first nonlinear medium. Now we retract all phase factors in $A(\bar{\omega}_p)$ and $e^{i\bar{\omega}_p \left|\tau_{0P} - \frac{\tau_{0A}}{2} - \frac{\tau_{0B}}{2} + \tau_{1P} - \frac{\tau_{1A}}{2} - \frac{\tau_{1B}}{2}\right|}$ and rewrite **Equation 60** as:

$$\left\langle \tilde{B} \right\rangle_{avg} = \gamma \int d\omega_P d\omega_A d\omega_B \left| \alpha \mathfrak{A}_0^{(type\,II)} L_1 sinc(\Delta k^{(1)}_{VHV} L_1/2) f(\omega_P) \right|^2 \times$$
$$\left[ 1 + |U_{4P}|^2 + 2|U_{4P}| e^{-\frac{|\Delta\tau_0 + \Delta\tau_1|}{\tau_C}} cos\left(\Delta k^{(0)} L_0 + \Delta k^{(1)}_{VHV} L_1 + \phi_P\right) \right] \delta(\omega_P - \omega_A - \omega_B), \quad (67)$$

where $\Delta\tau_0 = \tau_{0P} - \frac{\tau_{0A}}{2} - \frac{\tau_{0B}}{2}$ and $\Delta\tau_1 = \tau_{1P} - \frac{\tau_{1A}}{2} - \frac{\tau_{1B}}{2}$; Note that if $|\Delta\tau_0 + \Delta\tau_1| \ll \tau_C$, the spectral interference occurs in the spectrum of the biphoton (integrand in **Equation 67**), which depends on the dispersion of the middle section. As can be seen from **Equation 67**, the integrand of $\left\langle \tilde{B} \right\rangle_{avg}$ is proportional to the integrand of the total biphoton brightness $\langle B_{tot} \rangle_{avg}$ defined in the paper and reveals the emission spectrum of the biphoton generated in the cascade structure. This suggests that by coincidence detection, one can obtain the spectrum of the biphotons generated in the cascade structure.

# 9 Biphoton polarization state in the cascade structure

In this section, we study the polarization state of the biphotons generated from the cascade structure and characterize the degree of polarization entanglement, by considering the concurrence[9, 10], for various polarization transformations in the middle section.

## 9.1 Polarization entanglement in the cascade structure:

The unitary transformation of the middle section is of the form mentioned in the paper[$\hat{U}_n(\theta, \phi_1, \phi_2) = \begin{pmatrix} e^{i\phi_1}\cos\theta & -e^{-i\phi_2}\sin\theta \\ e^{i\phi_2}\sin\theta & e^{i\phi_1}\cos\theta \end{pmatrix}$]. Here we first consider the biphoton polarization state for type II SPDC cascade when there is no polarization rotation in the middle section [$\hat{U}_n(\theta = \phi_1 = \phi_2 = 0)$]. The normalized density matrix ($\rho = \frac{|\psi_{out}^{typeII}\rangle\langle\psi_{out}^{typeII}|}{\langle\psi_{out}^{typeII}|\psi_{out}^{typeII}\rangle}$) for this case can be expressed as:

$$\rho_{(\theta=\phi_1=\phi_2=0)}^{typeII} = \frac{1}{k}\sum_{\mathcal{L},\mathcal{M}}\alpha_{\mathcal{L}}\alpha_{\mathcal{M}}^*\int d\omega_A d\omega_B \begin{pmatrix} 0 & 0 & 0 & 0 \\ 0 & \phi_{\mathcal{L},HV}(\omega_A,\omega_B)\phi_{\mathcal{M},HV}^*(\omega_A,\omega_B) & \phi_{\mathcal{L},HV}(\omega_A,\omega_B)\phi_{\mathcal{M},VH}^*(\omega_A,\omega_B) & 0 \\ 0 & \phi_{\mathcal{L},VH}(\omega_A,\omega_B)\phi_{\mathcal{M},HV}^*(\omega_A,\omega_B) & \phi_{\mathcal{L},VH}(\omega_A,\omega_B)\phi_{\mathcal{M},VH}^*(\omega_A,\omega_B) & 0 \\ 0 & 0 & 0 & 0 \end{pmatrix},$$
(68)

where $k = \sum_{S'S''}\int d\omega_A d\omega_B \left\langle \sum_{\mathcal{L}}\left|\alpha_{\mathcal{L}}\phi_{\mathcal{L},S'S''}(\omega_A,\omega_B)\right|^2 \right\rangle_{avg}$; $\phi_{\mathcal{L},HV}(\omega_A,\omega_B)$ and $\phi_{\mathcal{L},VH}(\omega_A,\omega_B)$ can be derived from **Equation 35**. Since for the given polarization rotation, $U_{2A} = U_{2B} = U_{3A} = U_{3B} = 0$ and $U_{4P} = 1$, we can write **Equation 68** in a simple form of

$$\rho_{(\theta=\phi_1=\phi_2=0)}^{typeII} = \frac{1}{k}\begin{pmatrix} 0 & 0 & 0 & 0 \\ 0 & \rho_{22} & \rho_{23} & 0 \\ 0 & \rho_{32} & \rho_{33} & 0 \\ 0 & 0 & 0 & 0 \end{pmatrix},$$
(69)

where

$$\rho_{22} = \sum_{\mathcal{L},\mathcal{M}}\alpha_{\mathcal{L}}\alpha_{\mathcal{M}}^*\int d\omega_P d\omega_A d\omega_B \left|\mathfrak{A}_0^{(typeII)}L_1 sinc(\Delta k_{VHV}^{(1)}L_1/2)\left[1 + e^{i\Delta k^{(0)}L_0}e^{i\Delta k_{VHV}^{(1)}L_1}\right]\right|^2 f_{\mathcal{L}}(\omega_P)f_{\mathcal{M}}^*(\omega_P)\delta(\omega_P - \omega_A - \omega_B),$$

$$\rho_{33} = \sum_{\mathcal{L},\mathcal{M}}\alpha_{\mathcal{L}}\alpha_{\mathcal{M}}^*\int d\omega_P d\omega_A d\omega_B \left|\mathfrak{A}_0^{(typeII)}L_1 sinc(\Delta k_{VHV}^{(1)}L_1/2)e^{2i\Lambda}\left[e^{i\Lambda} + e^{-i\Lambda}e^{i\Delta k^{(0)}L_0}e^{i\Delta k_{VHV}^{(1)}L_1}\right]\right|^2 f_{\mathcal{L}}(\omega_P)f_{\mathcal{M}}^*(\omega_P)\delta(\omega_P - \omega_A - \omega_B),$$

and for $\rho_{23}$ and $\rho_{32}$ we have:

$$\rho_{23} = \rho_{32}^* = \sum_{\mathcal{L},\mathcal{M}}\alpha_{\mathcal{L}}\alpha_{\mathcal{M}}^*\int d\omega_P d\omega_A d\omega_B \left|\mathfrak{A}_0^{(typeII)}L_1 sinc(\Delta k_{VHV}^{(1)}L_1/2)\right|^2 f_{\mathcal{L}}(\omega_P)f_{\mathcal{M}}^*(\omega_P)\delta(\omega_P - \omega_A - \omega_B)\times$$

$$e^{-2i\Lambda}\left[1 + e^{i\Delta k^{(0)}L_0}e^{i\Delta k_{VHV}^{(1)}L_1}\right]\left[e^{i\Lambda} + e^{-i\Lambda}e^{i\Delta k^{(0)}L_0}e^{i\Delta k_{VHV}^{(1)}L_1}\right]^*. \quad (70)$$

Since for PPSF we have $\Lambda \ll \Delta k^{(0)} L_0 + \Delta k_{VHV}^{(1)} L_1$ [7], $cos(\Delta k^{(0)} L_0 + \Delta k_{VHV}^{(1)} L_1) \approx cos(\Delta k^{(0)} L_0 + \Delta k_{VHV}^{(1)} L_1 - 2\Lambda)$ and we can write the ensemble average of the density matrix in **Equation 69** as:

$$\left\langle \rho_{(\theta=\phi_1=\phi_2=0)}^{type\,II} \right\rangle_{avg} = \frac{1}{k} \int d\omega_P d\omega_A d\omega_B \left| \alpha \mathfrak{A}_0^{(typeII)} L_1 sinc(\Delta k_{VHV}^{(1)} L_1/2) f(\omega_P) \right|^2 \delta(\omega_P - \omega_A - \omega_B) \times$$

$$\left( 2 + 2e^{-\frac{|\Delta\tau_0+\Delta\tau_1|}{\tau_C}} cos(\Delta k^{(0)} L_0 + \Delta k_{VHV}^{(1)} L_1) \right) \begin{pmatrix} 0 & 0 & 0 & 0 \\ 0 & 1 & e^{-2i\Lambda} & 0 \\ 0 & e^{2i\Lambda} & 1 & 0 \\ 0 & 0 & 0 & 0 \end{pmatrix}. \quad (71)$$

In the case of incoherent pumping ($\tau_C \ll |\Delta\tau_0 + \Delta\tau_1|$), the interference term that include $e^{-\frac{|\Delta\tau_0+\Delta\tau_1|}{\tau_C}}$ vanishes, and the density matrix becomes:

$$\left\langle \rho_{(\theta=\phi_1=\phi_2=0)}^{type\,II} \right\rangle_{avg} = \frac{2}{k} \int d\omega_P d\omega_A d\omega_B \left| \alpha \mathfrak{A}_0^{(typeII)} L_1 sinc(\Delta k_{VHV}^{(1)} L_1/2) f(\omega_P) \right|^2 \delta(\omega_P - \omega_A - \omega_B) \times \begin{pmatrix} 0 & 0 & 0 & 0 \\ 0 & 1 & e^{-2i\Lambda} & 0 \\ 0 & e^{2i\Lambda} & 1 & 0 \\ 0 & 0 & 0 & 0 \end{pmatrix}, \quad (72)$$

To have maximum concurrence of 1, the off-diagonal elements should not disappear, which requires

$$\int d\omega_A d\omega_B e^{-2i\Lambda} \left| L_1 sinc(\Delta k_{VHV}^{(1)} L_1/2) \right|^2 \neq 0.$$

Since the function $L_1 sinc(\Delta k_{VHV}^{(1)} L_1/2)$ is slowly-varying compare to $e^{-2i\Lambda}$, the criteria of having nonzero off-diagonal elements will reduce to $\Lambda \ll 1$ for all signal and idler frequencies. This criteria is only met in materials with negligible group birefringence [7]. Note that the results mentioned above are also valid for the *coherent pumping* as well.

### 9.2 Biphoton state with polarization rotation in the middle section:

Now we study the concurrence when the polarization transformation in the middle section is $\hat{U}_n(\theta = \frac{\pi}{4}, \phi_1 = \phi_2 = 0)$. For PPSF $\Lambda \ll 1$, so we replace $e^{i\Lambda}$ with 1. After the ensemble averaging, we can write the density as below:

$$\rho_{(\theta=\frac{\pi}{4},\phi_1=\phi_2=0)}^{type\,II} = \frac{1}{k} \int d\omega_P d\omega_A d\omega_B \left| \alpha \mathfrak{A}_0^{(typeII)} L_1 sinc(\Delta k_{VHV}^{(1)} L_1/2) f(\omega_P) \right|^2 \delta(\omega_P - \omega_A - \omega_B) \times$$

$$\begin{pmatrix} 1 & -e^{-\frac{|\Delta\tau_0+\Delta\tau_1|}{\tau_C}} \rho_1^* & -e^{-\frac{|\Delta\tau_0+\Delta\tau_1|}{\tau_C}} \rho_1^* & -1 \\ -e^{-\frac{|\Delta\tau_0+\Delta\tau_1|}{\tau_C}} \rho_1 & 1 & 1 & e^{-\frac{|\Delta\tau_0+\Delta\tau_1|}{\tau_C}} \rho_2 \\ -e^{-\frac{|\Delta\tau_0+\Delta\tau_1|}{\tau_C}} \rho_1 & 1 & 1 & e^{-\frac{|\Delta\tau_0+\Delta\tau_1|}{\tau_C}} \rho_2 \\ -1 & e^{-\frac{|\Delta\tau_0+\Delta\tau_1|}{\tau_C}} \rho_2^* & e^{-\frac{|\Delta\tau_0+\Delta\tau_1|}{\tau_C}} \rho_2^* & 1 \end{pmatrix}, \quad (73)$$

where $\rho_1 = |U_{4P}| e^{i\Delta k^{(0)} L_0} e^{i\Delta k_{VHV}^{(1)} L_1} e^{-i\Gamma_B}$ and $\rho_2 = |U_{4P}| e^{i\Delta k^{(0)} L_0} e^{i\Delta k_{VHV}^{(1)} L_1} e^{-i\Gamma_A}$. Note that the density matrix in coherent regime ($|\Delta\tau_0 + \Delta\tau_1| \ll \tau_C$) is frequency-dependent. In fact, one can perform a simulation and carry out the concurrence as a function of the signal and idler frequencies for various biphoton sources. This was done for PPSF and the result can be found in **Fig. 4a** in the paper. In the limit when $\tau_C \ll |\Delta\tau_0 + \Delta\tau_1|$ (incoherent pumping) we find that

$$\left\langle \rho_{(\theta=\frac{\pi}{4},\phi_1=\phi_2=0)}^{type\,II} \right\rangle_{avg} = \frac{1}{4} \begin{pmatrix} 1 & 0 & 0 & -1 \\ 0 & 1 & 1 & 0 \\ 0 & 1 & 1 & 0 \\ -1 & 0 & 0 & 1 \end{pmatrix}, \quad (74)$$

which corresponds to the biphoton polarization state with zero concurrence.